\RequirePackage{fix-cm}
\documentclass[12pt]{article}

\newif\ifAnonymous
\Anonymousfalse

\usepackage[margin=0.9in]{geometry}
\usepackage[T1]{fontenc}
\usepackage[utf8]{inputenc}
\usepackage[scaled=0.95]{helvet}

\usepackage{mathtools}
\mathtoolsset{showonlyrefs}
\usepackage{amssymb}
\usepackage{amsthm}
\usepackage{newtxmath}
\usepackage{booktabs}
\usepackage{enumitem}
\usepackage{microtype}
\usepackage{bm}
\usepackage[numbers,square,sort&compress]{natbib}
\bibpunct[, ]{[}{]}{,}{n}{,}{,}
\makeatletter
\def\NAT@def@citea{\def\@citea{\NAT@separator}}
\makeatother
\usepackage[dvipsnames]{xcolor}
\usepackage{tabularx}
\usepackage{graphicx}
\usepackage{caption}
\usepackage{subcaption}
\usepackage{float}
\usepackage{placeins}
\usepackage{lscape}
\usepackage{multirow}
\usepackage{ragged2e}
\usepackage{titlesec}

\usepackage{longtable}

\titleformat{\section}{\normalfont\large\bfseries}{\thesection.}{0.6em}{}
\titleformat{\subsection}{\normalfont\large\bfseries}{\thesubsection.}{0.6em}{}
\titleformat{\subsubsection}{\normalfont\normalsize\bfseries}{\thesubsubsection.}{0.6em}{}

\graphicspath{{Plots/}}

\floatstyle{plaintop}
\newfloat{algorithm}{tbp}{loa}
\floatname{algorithm}{Algorithm}

\usepackage{hyperref}
\hypersetup{
  colorlinks = true,
  linkcolor = blue,
  citecolor = blue,
  urlcolor = blue,
  hypertexnames = false
}

\definecolor{EquationRed}{HTML}{8B2635}

\let\OriginalEqref\eqref
\renewcommand{\eqref}[1]{%
  \begingroup
  \hypersetup{linkcolor=EquationRed}%
  \OriginalEqref{#1}%
  \endgroup
}

\usepackage{doi}
\usepackage{orcidlink}

\usepackage{afterpage}

\renewcommand{\sectionautorefname}{Section}
\renewcommand{\subsectionautorefname}{Section}
\renewcommand{\subsubsectionautorefname}{Section}
\providecommand{\algorithmautorefname}{Algorithm}

\newcommand{\mainsecref}[1]{\hyperref[#1]{Section~\ref*{#1}}}
\newcommand{\mainfigref}[1]{\hyperref[#1]{Figure~\ref*{#1}}}
\newcommand{\maintabref}[1]{\hyperref[#1]{Table~\ref*{#1}}}
\newcommand{\mainalgref}[1]{\hyperref[#1]{Algorithm~\ref*{#1}}}

\newcommand{\appref}[1]{\hyperref[#1]{Appendix~\ref*{#1}}}
\newcommand{\apptabref}[1]{\hyperref[#1]{Appendix Table~\ref*{#1}}}
\newcommand{\appfigref}[1]{\hyperref[#1]{Appendix Figure~\ref*{#1}}}
\newcommand{\appalgref}[1]{\hyperref[#1]{Appendix Algorithm~\ref*{#1}}}

\newcommand{\suppsecref}[1]{\hyperref[#1]{Section~\ref*{#1}}}
\newcommand{\supptabref}[1]{\hyperref[#1]{Table~\ref*{#1}}}
\newcommand{\suppfigref}[1]{\hyperref[#1]{Figure~\ref*{#1}}}
\newcommand{\suppalgref}[1]{\hyperref[#1]{Algorithm~\ref*{#1}}}
\newcommand{\suppeqref}[1]{\hyperref[#1]{Equation~\ref*{#1}}}

\newcommand{\E}{\mathbb{E}}
\newcommand{\Var}{\operatorname{Var}}

\newcommand{\Normal}{\mathcal{N}}

\newcommand{\R}{\mathbb{R}}
\newcommand{\obs}{\mathcal{O}}

\newcommand{\DK}{\operatorname{DK}}
\newcommand{\LKJ}{\operatorname{LKJ}}
\newcommand{\HalfNormal}{\operatorname{HalfNormal}}
\newcommand{\Uniform}{\operatorname{Uniform}}
\DeclareMathOperator{\logit}{logit}
\DeclareMathOperator{\expit}{expit}
\DeclareMathOperator{\diag}{diag}

\theoremstyle{plain}
\newtheorem{proposition}{Proposition}

\newcommand{\ManuscriptTitle}{A Bayesian quantile mixed model for bounded discrete outcomes}
\definecolor{ReferenceBlue}{HTML}{000080}

\hypersetup{
  pdftitle = {A Bayesian quantile mixed model for bounded discrete outcomes},
  colorlinks = true,
  linkcolor = ReferenceBlue,
  citecolor = ReferenceBlue,
  urlcolor = ReferenceBlue
}

\ifAnonymous
  \hypersetup{pdfauthor = {}}
\else
  \hypersetup{pdfauthor = {Divan Aristo Burger}}
\fi

\newcommand{\AuthorBlock}{
Divan Aristo Burger\,\orcidlink{0000-0001-8096-6371}$^{1,2,*}$
}

\newcommand{\AffiliationBlock}{
\begin{tabularx}{\textwidth}{@{}r@{\hspace{0.4em}}X@{}}
$^{1}$ & Cytel Inc., Waltham, MA, USA\\
$^{2}$ & Department of Mathematical Statistics and Actuarial Science, University of the Free State, Bloemfontein, South Africa
\end{tabularx}
}

\newcommand{\CorrespondingAuthor}{
\begin{tabularx}{\textwidth}{@{}r@{\hspace{0.27em}}X@{}}
$^{*}$ & Corresponding author: Divan Aristo Burger, Cytel Inc., 1050 Winter Street, Waltham, MA 02451, USA. Email: \href{mailto:divanaburger@gmail.com}{\texttt{divanaburger@gmail.com}}
\end{tabularx}
}

\newcommand{\MainAbstractText}{
Quantile regression for longitudinal bounded integer outcomes must accommodate discreteness, finite support, and dependence within participants. We develop a Bayesian quantile mixed model using a discrete Kumaraswamy distribution parameterized through a smooth quantile anchor whose ceiling gives the corresponding integer quantile. Covariates and correlated Gaussian participant effects enter a logit predictor for the scaled anchor, while a separate regression models distributional shape. Posterior computation combines a conjugate Gaussian coefficient update, available for compatible fixed and random effect designs, with slice sampling and covariance interweaving. Simulations showed small biases in regression coefficients and generally near-nominal coverage of 95\% credible intervals. An application to repeated cognitive scores illustrates covariate associations across separately fitted quantile levels. Posterior predictive checks captured broad distributional patterns.
}

\newcommand{\ManuscriptWordCount}{5739}

\newcommand{\ManuscriptMetadataBlock}{%
\noindent
\begin{tabularx}{\textwidth}{@{}l@{\hspace{0.35em}}>{\justifying\setlength{\parindent}{0pt}\arraybackslash}X@{}}
\textbf{Keywords:} &
Bayesian inference; Bounded discrete outcomes; Kumaraswamy distribution; Longitudinal data; Quantile regression; Random effects
\end{tabularx}%
\par\vspace{0.2cm}
\noindent\mbox{\textbf{Word count:}\hspace{0.35em}\ManuscriptWordCount}%
\par
}

\newcommand{\MainTitlePageContent}{
\noindent{\color{ReferenceBlue} \{We develop a Bayesian quantile mixed model for bounded longitudinal scores. Posterior computation combines a conjugate Gaussian coefficient update, available when the fixed and random effect designs are compatible, with elliptical slice sampling and covariance interweaving. The sampler is implemented in C++, with R coordinating fitting and summaries.\}}

\medskip

\noindent\textbf{Abstract}

\vspace{0.25cm}

\noindent\ignorespaces\MainAbstractText

\vspace{0.4cm}

\ManuscriptMetadataBlock
}

\newcommand{\SupplementTitle}{
Supplementary material for:\\[0.35em]
\ManuscriptTitle
}

\newcommand{\IfNotEmpty}[2]{
\if\relax\detokenize{#1}\relax
\else
#2
\fi
}

\newcommand{\MakeManuscriptTitlePage}[2]{
\begin{titlepage}
\thispagestyle{empty}

{\large\bfseries #1\par}

\vspace{0.7cm}

\rule{\textwidth}{0.4pt}

\vspace{0.7cm}

\ifAnonymous
{\large Author details withheld for anonymous review.\par}
\else
{\normalsize \AuthorBlock\par}

\vspace{0.55cm}

\begin{minipage}{\textwidth}
{\small \AffiliationBlock\par}
\end{minipage}

\vspace{0.45cm}

\begin{minipage}{\textwidth}
{\normalsize \CorrespondingAuthor\par}
\end{minipage}

\fi

\vspace{0.7cm}

\IfNotEmpty{#2}{
\begin{minipage}{\textwidth}
#2
\end{minipage}
}

\end{titlepage}
\newpage
}

\newcommand{\StartSupplement}{
\clearpage
\titleformat{\section}{\normalfont\large\bfseries}{\thesection.}{0.6em}{}
\MakeManuscriptTitlePage{\SupplementTitle}{}
\pagenumbering{arabic}
\pagestyle{plain}
\setcounter{section}{0}
\setcounter{subsection}{0}
\setcounter{subsubsection}{0}
\setcounter{figure}{0}
\setcounter{table}{0}
\setcounter{algorithm}{0}
\setcounter{equation}{0}
\renewcommand{\thesection}{S\arabic{section}}
\renewcommand{\thesubsection}{S\arabic{section}.\arabic{subsection}}
\renewcommand{\thesubsubsection}{S\arabic{section}.\arabic{subsection}.\arabic{subsubsection}}
\renewcommand{\thefigure}{S\arabic{figure}}
\renewcommand{\thetable}{S\arabic{table}}
\renewcommand{\thealgorithm}{S\arabic{algorithm}}
\renewcommand{\theequation}{S\arabic{equation}}
\renewcommand{\theHsection}{supp.\arabic{section}}
\renewcommand{\theHsubsection}{supp.\arabic{section}.\arabic{subsection}}
\renewcommand{\theHsubsubsection}{supp.\arabic{section}.\arabic{subsection}.\arabic{subsubsection}}
\renewcommand{\theHfigure}{supp.\arabic{figure}}
\renewcommand{\theHtable}{supp.\arabic{table}}
\providecommand{\theHalgorithm}{\thealgorithm}
\renewcommand{\theHalgorithm}{supp.\arabic{algorithm}}
\renewcommand{\theHequation}{supp.\arabic{equation}}
\renewcommand{\sectionautorefname}{Section}
\renewcommand{\subsectionautorefname}{Section}
\renewcommand{\subsubsectionautorefname}{Section}
\renewcommand{\figureautorefname}{Figure}
\renewcommand{\tableautorefname}{Table}
\renewcommand{\equationautorefname}{Equation}
\renewcommand{\algorithmautorefname}{Algorithm}
\captionsetup[figure]{name = Figure}
\captionsetup[table]{name = Table}
}

\begin{document}

\MakeManuscriptTitlePage{\ManuscriptTitle}{\MainTitlePageContent}

\section{Introduction}
\label{sec:introduction}

Bounded discrete outcomes occur when an assessment takes values on a finite integer scale or a count has a known maximum. Examples include cognitive test scores, symptom totals, and the number of successes among a fixed number of opportunities. Repeated measurements introduce dependence within participants, while skewness and concentrations near the boundaries can make changes in the outcome distribution difficult to summarize by changes in its mean alone. Quantile regression provides a way to describe covariate associations at different locations of that distribution \citep{koenker2017quantile}.

For discrete outcomes, quantile regression requires care in defining the regression target. The usual quantile changes in steps, and a probability level need not be attained exactly by the cumulative distribution function (CDF). Existing longitudinal approaches include Bayesian regression for smoothed counts \citep{jantre2023bayesian} and conditional mid-quantile regression \citep{russo2024mid}. A continuous extension can provide a smooth target while retaining a discrete observation model, as in model-aware quantile regression \citep{padellini2018model}. Related median regressions use a smooth pseudo-median for binomial-logit-$t$ outcomes \citep{burger2026outliers} and a shifted median for contaminated discrete Weibull counts \citep{burger2026robust}.

For continuous bounded responses, Mitnik and Baek~\citep{mitnik2013kumaraswamy} developed median-based Kumaraswamy parameterizations, and Bayes et al.~\citep{bayes2017quantile} formulated Bayesian quantile mixed regression with predictors for both the quantile and shape parameters. For bounded integer outcomes, Barbiero and Hitaj~\citep{barbiero2026discrete} introduced the discrete Kumaraswamy distribution, with an explicit probability mass function (PMF) and quantile function, and regression models for its shape parameters. We build on these developments while distinguishing the smooth quantile anchor from the corresponding integer quantile.

The proposed model extends this discrete family to longitudinal quantile regression, combining a smooth quantile anchor, a shape regression, and correlated participant effects. We make the anchor-to-integer mapping explicit, distinguish conditional and marginal quantiles, and derive a Gaussian full conditional update for the fixed regression coefficients when the fixed and random effect designs satisfy a specified compatibility condition. The sampler combines this update with established slice sampling and interweaving methods \citep{neal2003slice,murray2010elliptical,yu2011interweaving}. Separate fits at several quantile levels describe covariate associations across the conditional distribution, subject to an assessment of crossing. Beta-binomial mixed models provide an established approach to modeling the conditional mean of longitudinal bounded integer outcomes \citep{najera2019beta}. Although quantiles can be derived from their fitted distributions, our formulation specifies the regression structure directly through a smooth quantile anchor.

\mainsecref{sec:distribution} gives the distributional construction and quantile relationships, \mainsecref{sec:mixed-model} specifies the mixed model, and \mainsecref{sec:computation} develops posterior computation. \mainsecref{sec:simulation} evaluates parameter recovery and interval coverage, and \mainsecref{sec:application} illustrates the method using repeated cognitive scores. \mainsecref{sec:discussion} discusses the findings, limitations, and possible extensions.

\section{The discrete Kumaraswamy distribution}
\label{sec:distribution}

\subsection{Construction and basic properties}
\label{sec:construction}
\label{sec:moments}

Let $V$ have the continuous Kumaraswamy distribution with positive shape parameters $a$ and $b$, and CDF
\begin{equation}
G\left(v \middle| a, b\right)
= 1 - \left(1 - v^a\right)^b,
\qquad 0 < v < 1.
\label{eq:continuous-cdf}
\end{equation}
For a known positive integer $K$, define $Y = \left\lfloor\left(K + 1\right)V\right\rfloor$. Then $Y$ takes values in $\mathcal{Y}_K = \left\{0, 1, \ldots, K\right\}$. This construction yields the distribution of Barbiero and Hitaj~\citep{barbiero2026discrete} shifted to start at zero, with $K + 1$ categories.
At integer arguments, the CDF and PMF are
\begin{equation}
F_K\left(y \middle| a, b\right)
= 1 - \left[1 - \left(\frac{y + 1}{K + 1}\right)^a\right]^b,
\qquad y \in \mathcal{Y}_K,
\label{eq:discrete-cdf}
\end{equation}
and
\begin{equation}
\begin{split}
p_K\left(y \middle| a, b\right)
&= F_K\left(y \middle| a, b\right)
 - F_K\left(y - 1 \middle| a, b\right)\\
&= \left[1 - \left(\frac{y}{K + 1}\right)^a\right]^b
 - \left[1 - \left(\frac{y + 1}{K + 1}\right)^a\right]^b,
\qquad y \in \mathcal{Y}_K.
\end{split}
\label{eq:pmf}
\end{equation}
The boundary conventions are $F_K\left(y \middle| a, b\right) = 0$ for $y < 0$ and $F_K\left(y \middle| a, b\right) = 1$ for $y \geq K$. At noninteger arguments in $\left[0, K\right)$, the actual discrete distribution function is obtained by replacing $y$ in \eqref{eq:discrete-cdf} by $\left\lfloor y\right\rfloor$. The probabilities in \eqref{eq:pmf} are positive and sum to one by telescoping.

The two boundary probabilities are
\begin{equation}
\Pr\left(Y = 0\right)
= 1 - \left[1 - \left(\frac{1}{K + 1}\right)^a\right]^b,
\qquad
\Pr\left(Y = K\right)
= \left[1 - \left(\frac{K}{K + 1}\right)^a\right]^b.
\label{eq:boundary-probabilities}
\end{equation}
Thus, the model assigns positive mass to both endpoints without an additional mixture component. These masses are determined by the same two shape parameters as the interior probabilities; they are not free parameters.

The upper bound $K$ is determined by the outcome scale. We require $K \geq 2$, corresponding to at least three possible outcome values, because a binary distribution cannot identify two unrestricted shape parameters. For an outcome supported on consecutive integers from a known lower bound $L$ to a known upper bound $U$, subtracting $L$ gives the support $\left\{0, 1, \ldots, K\right\}$ with $K = U - L$.

All moments exist because the support is finite. Writing
\begin{equation}
S_K\left(y \middle| a, b\right)
= \Pr\left(Y > y\right)
= \left[1 - \left(\frac{y + 1}{K + 1}\right)^a\right]^b,
\qquad y = 0, \ldots, K - 1,
\end{equation}
the first two moments follow from the survival-sum identities
\begin{equation}
\E\left(Y\right)
= \sum_{y=0}^{K-1} S_K\left(y \middle| a, b\right),
\qquad
\E\left(Y^2\right)
= \sum_{y=0}^{K-1} \left(2y + 1\right)S_K\left(y \middle| a, b\right).
\label{eq:moments}
\end{equation}
The variance is $\Var\left(Y\right) = \E\left(Y^2\right) - \left[\E\left(Y\right)\right]^2$.

Although the mean can be evaluated as a finite sum, expressing a shape parameter in terms of the mean while holding the other fixed generally requires numerical inversion. The smooth quantile anchor introduced below instead permits an explicit reparameterization.

\subsection{Smooth extension and integer quantiles}
\label{sec:quantiles}

Define a smooth extension of the integer-point distribution function by
\begin{equation}
H_K\left(s \middle| a, b\right)
= 1 - \left[1 - \left(\frac{s + 1}{K + 1}\right)^a\right]^b,
\qquad -1 < s < K,
\label{eq:smooth-cdf}
\end{equation}
with $H_K\left(s \middle| a, b\right) = 0$ for $s \leq -1$ and $H_K\left(s \middle| a, b\right) = 1$ for $s \geq K$. This is the distribution function of $W = \left(K + 1\right)V - 1$ and agrees with \eqref{eq:discrete-cdf} at the integer support points. It differs from the step function between those points. The construction gives $Y = \left\lceil W\right\rceil$ almost surely; no independently jittered response is introduced.

For $u \in \left(0, 1\right)$, the inverse of \eqref{eq:smooth-cdf} is
\begin{equation}
\xi_u
= \left(K + 1\right)
\left[1 - \left(1 - u\right)^{1/b}\right]^{1/a} - 1.
\label{eq:anchor-general}
\end{equation}
We call $\xi_u$ the smooth quantile anchor, or relaxed quantile. It is the $u$-quantile of the specified smooth extension, rather than a noninteger value of the discrete quantile function.

Proposition~\ref{prop:integer-quantile} makes explicit the relationship between the smooth quantile anchor and the integer quantile and recovers the quantile formula of Barbiero and Hitaj~\citep{barbiero2026discrete} under our shifted support.

\begin{proposition}
\label{prop:integer-quantile}
For $a, b > 0$, and $u \in \left(0, 1\right)$, the lower integer quantile satisfies
\begin{equation}
Q_Y\left(u\right)
:= \min\left\{y \in \mathcal{Y}_K :
F_K\left(y \middle| a, b\right) \geq u\right\}
= \left\lceil \xi_u\right\rceil.
\label{eq:integer-quantile}
\end{equation}
In particular, $0 \leq Q_Y\left(u\right) - \xi_u < 1$.
\end{proposition}

\begin{proof}
The function $H_K$ is strictly increasing on $\left(-1, K\right)$ and $H_K\left(\xi_u \middle| a, b\right) = u$. At an integer $y \in \mathcal{Y}_K$, $F_K\left(y \middle| a, b\right) \geq u$ is therefore equivalent to $y \geq \xi_u$. Since $-1 < \xi_u < K$, the smallest such integer is $\left\lceil \xi_u\right\rceil$. The bound follows from the defining property of the ceiling function.
\end{proof}

In practical terms, the discrete quantile is the relaxed quantile rounded upward to the nearest integer: a relaxed quantile of 7.3 gives an integer quantile of 8, whereas a relaxed quantile of 7 gives 7. Thus, the discrete quantile lies less than one unit above the relaxed quantile. This relationship also gives an exact simulation rule: if $U^\star \sim \Uniform\left(0, 1\right)$, then $Y = \left\lceil \xi_{U^\star}\right\rceil$ has PMF \eqref{eq:pmf}.

\subsection{Quantile parameterization}
\label{sec:parameterization}

Fix a target level $\tau \in \left(0, 1\right)$. Define the scaled version of the anchor in \eqref{eq:anchor-general} by
\begin{equation}
r_\tau = \frac{\xi_\tau + 1}{K + 1},
\qquad 0 < r_\tau < 1.
\label{eq:scaled-anchor}
\end{equation}
For a positive shape parameter $\phi$, set
\begin{equation}
a = -\frac{\phi}{\log\left(r_\tau\right)},
\qquad
b = \frac{\log\left(1 - \tau\right)}
{\log\left(1 - \exp\left(-\phi\right)\right)}.
\label{eq:quantile-parameterization}
\end{equation}
Both shapes are positive throughout the rectangular parameter space $\left(r_\tau, \phi\right) \in \left(0, 1\right) \times \left(0, \infty\right)$. Equivalently, $-1 < \xi_\tau < K$ and $\phi > 0$. Since $r_\tau^a = \exp\left(-\phi\right)$, substitution into \eqref{eq:smooth-cdf} gives
\begin{equation}
H_K\left(\xi_\tau \middle| a, b\right)
= 1 - \left[1 - \exp\left(-\phi\right)\right]^b
= \tau.
\label{eq:anchor-property}
\end{equation}
Consequently, \eqref{eq:quantile-parameterization} preserves the chosen relaxed quantile exactly, and the corresponding integer quantile is $\left\lceil \xi_\tau\right\rceil$.

For fixed $\tau$, this is a reparameterization of the original two-shape family. Given $a, b > 0$, the inverse mapping is
\begin{equation}
\phi = -\log\left[1 - \left(1 - \tau\right)^{1/b}\right],
\qquad
r_\tau = \left[1 - \left(1 - \tau\right)^{1/b}\right]^{1/a}.
\label{eq:inverse-parameterization}
\end{equation}
Here, $\tau$ indexes the chosen regression target and is not an additional unknown distributional parameter. We write $Y \sim \DK_K\left(r_\tau, \phi; \tau\right)$ for the PMF in \eqref{eq:pmf} after this substitution.

For the continuous variable $V$, $r_\tau$ is its ordinary $\tau$-quantile, and \eqref{eq:quantile-parameterization} is the parameterization of Bayes et al.~\citep{bayes2017quantile}, with their $\kappa$ corresponding to $r_\tau$. Here, this parameterization indexes the discrete probabilities in \eqref{eq:pmf}, with relaxed anchor $\xi_\tau = \left(K + 1\right)r_\tau - 1$ and integer quantile $\left\lceil \xi_\tau \right\rceil$.

For fixed $\tau$ and $r_\tau$, the parameter $\phi$ changes the distributional shape while leaving $\xi_\tau$ fixed. It is not defined as a variance or an inverse variance. The moments in \eqref{eq:moments} generally depend on both $r_\tau$ and $\phi$; neither the anchor nor the scaled anchor is parameterized as the response mean. Moreover, the smooth anchor is specific to the extension in \eqref{eq:smooth-cdf}. Other interpolations of the same discrete probabilities need not give the same noninteger anchor, even though the integer quantiles are unchanged.

\begin{figure}[!tbp]
\centering
\includegraphics[width=\textwidth]{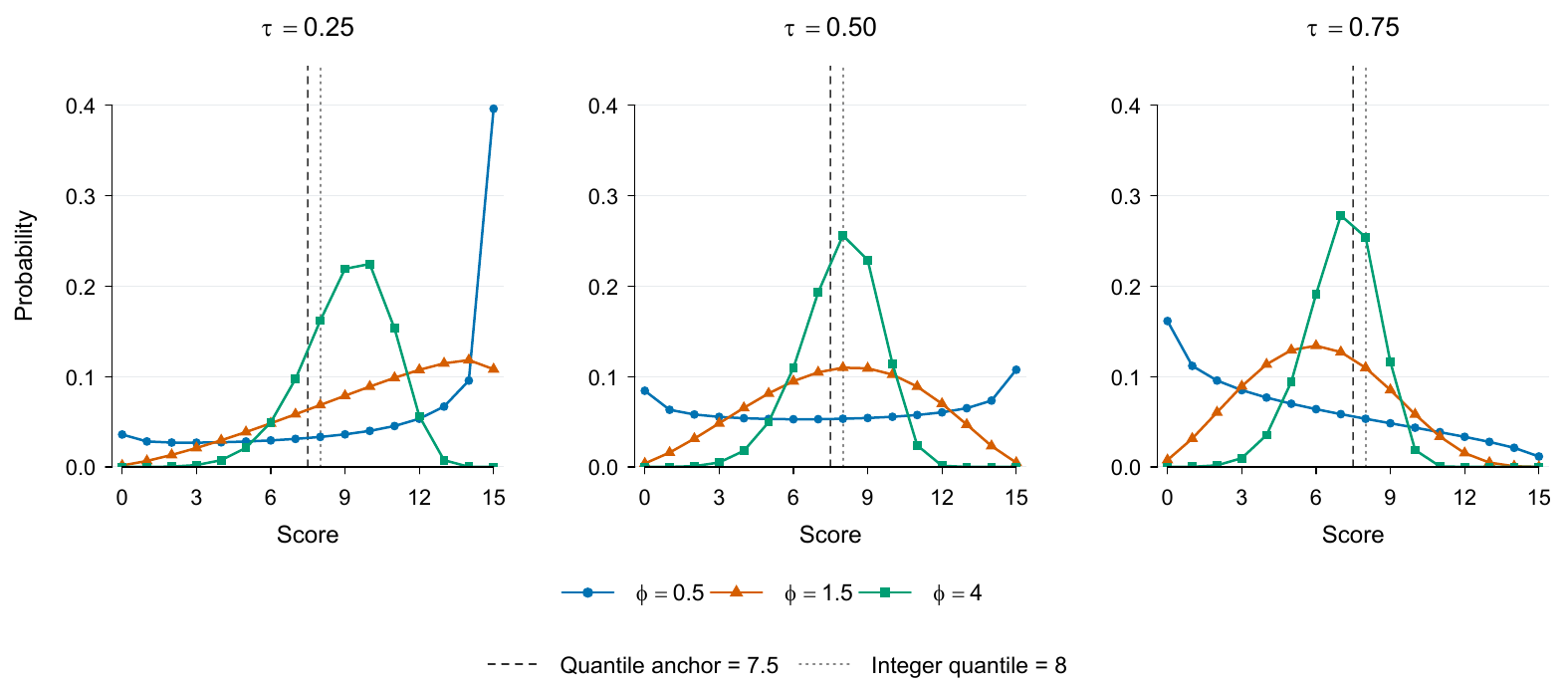}
\caption{Discrete Kumaraswamy probability masses on $\left\{0, \ldots, 15\right\}$ for $\phi \in \left\{0.5, 1.5, 4\right\}$ at three target quantile levels. All distributions share the relaxed anchor $\xi_\tau = 7.5$ and corresponding integer quantile $Q_Y\left(\tau\right) = 8$, marked by the dashed and dotted vertical lines, respectively. Points represent probability masses; connecting lines are visual guides.}
\label{fig:shape-illustration}
\end{figure}

\autoref{fig:shape-illustration} illustrates the role of $\phi$ at three target quantile levels. Within each panel, changing $\phi$ alters the probability masses while preserving the relaxed anchor at $\xi_\tau = 7.5$ and the corresponding integer quantile at $Q_Y\left(\tau\right) = 8$. The examples show that markedly different distributions can share the same target quantile. The anchor satisfies $H_K\left(7.5\right) = \tau$, whereas the integer quantile satisfies $F_K\left(7\right) < \tau < F_K\left(8\right)$; its cumulative probability need not equal $\tau$.

\section{Mixed quantile regression}
\label{sec:mixed-model}

\subsection{Model specification and interpretation}
\label{sec:general-specification}

Let $Y_{ij} \in \left\{0, 1, \ldots, K_{ij}\right\}$ denote the outcome for participant $i = 1, \ldots, n$ at assessment $j = 1, \ldots, m_i$, where $n$ is the number of participants, $m_i$ is the number of assessment records for participant $i$, and each upper bound $K_{ij} \geq 2$ is known. A common score maximum is the special case $K_{ij} = K$. Let $\bm{x}_{ij} \in \R^p$ be the fixed effect design vector, $\bm{z}_{ij} \in \R^q$ the random effect design vector, and $\bm{w}_{ij} \in \R^d$ the shape design vector. These vectors may contain intercepts, time functions, participant characteristics and interactions as appropriate. We condition throughout on the covariates, bounds and assessment times.

We write $\Normal_k\left(\bm{\mu},\bm{V}\right)$ for the $k$-variate normal distribution with mean vector $\bm{\mu}$ and covariance matrix $\bm{V}$, and $\varphi_k$ for its density. The symbols $\bm{0}_k$ and $\bm{I}_k$ denote the zero vector of length $k$ and the $k \times k$ identity matrix; the dimension of a zero vector is omitted when clear from context.

For a fixed $\tau$, specify
\begin{equation}
\begin{aligned}
\logit\left(r_{ij,\tau}\right)
&= \eta_{ij,\tau}
= \bm{x}_{ij}^{\mathsf T}\bm{\beta}_\tau
 + \bm{z}_{ij}^{\mathsf T}\bm{b}_{i,\tau},\\
\log\left(\phi_{ij,\tau}\right)
&= \bm{w}_{ij}^{\mathsf T}\bm{\gamma}_\tau,\\
\bm{b}_{i,\tau}
&\overset{\mathrm{ind}}{\sim}
\Normal_q\left(\bm{0}, \bm{\Sigma}_\tau\right),
\qquad i = 1, \ldots, n.
\end{aligned}
\label{eq:mixed-predictor}
\end{equation}
Here, $r_{ij,\tau}$ and $\phi_{ij,\tau}$ are the observation-specific scaled anchor and shape parameter, $\bm{\beta}_\tau \in \R^p$ and $\bm{\gamma}_\tau \in \R^d$ are their regression coefficient vectors, and $\eta_{ij,\tau}$ is the quantile linear predictor. The participant effects $\bm{b}_{i,\tau} \in \R^q$ have positive definite covariance matrix $\bm{\Sigma}_\tau$, and $\logit\left(r\right) = \log\left\{r/\left(1 - r\right)\right\}$. Conditional on $\bm{b}_{i,\tau}$ and the model parameters, the observations for participant $i$ are independent with
\begin{equation}
\left. Y_{ij} \middle| \bm{b}_{i,\tau}, \bm{\theta}_\tau \right.
\sim \DK_{K_{ij}}\left(r_{ij,\tau}, \phi_{ij,\tau}; \tau\right),
\label{eq:conditional-model}
\end{equation}
where $\bm{\theta}_\tau$ collects the regression and covariance parameters. In \eqref{eq:conditional-model}, the two Kumaraswamy shapes are
\begin{equation}
a_{ij,\tau} = -\frac{\phi_{ij,\tau}}{\log\left(r_{ij,\tau}\right)},
\qquad
b_{ij,\tau} = \frac{\log\left(1 - \tau\right)}
{\log\left(1 - \exp\left(-\phi_{ij,\tau}\right)\right)}.
\label{eq:conditional-shapes}
\end{equation}
The scalar $b_{ij,\tau}$ in \eqref{eq:conditional-shapes} is a distributional shape; the bold vector $\bm{b}_{i,\tau}$ denotes participant effects.

The conditional anchor and integer quantile are therefore
\begin{equation}
\begin{aligned}
\xi_{ij,\tau}
&= \left(K_{ij} + 1\right)\expit\left(\eta_{ij,\tau}\right) - 1,\\
Q_{ij}\left(\tau \middle| \bm{b}_{i,\tau}, \bm{\theta}_\tau\right)
&= \left\lceil \xi_{ij,\tau}\right\rceil,
\end{aligned}
\label{eq:conditional-quantile}
\end{equation}
where $\expit\left(\eta\right) = \left[1 + \exp\left(-\eta\right)\right]^{-1}$. Every finite linear predictor gives $-1 < \xi_{ij,\tau} < K_{ij}$, so \eqref{eq:conditional-quantile} respects the known integer support. Values of the relaxed anchor below zero correspond to an integer quantile of zero.

The design $\bm{z}_{ij}$ determines the participant-specific variation. A random intercept uses $\bm{z}_{ij} = 1$; an intercept and slope use $\bm{z}_{ij} = \left(1, t_{ij}\right)^{\mathsf T}$, where $t_{ij}$ is assessment time on a specified scale; and more general designs can include several slopes or prespecified time basis functions. Actual assessment times may differ across participants. Conditional independence in \eqref{eq:conditional-model} attributes within-participant dependence to these shared effects; an additional serial dependence process is not implicit in the formulation. Estimability of a chosen random effect structure depends on the number and timing of repeated measurements.

The coefficients $\bm{\beta}_\tau$ act on the logit of the scaled quantile anchor. For a scalar covariate that enters the fixed predictor linearly, a one-unit increase changes that predictor by its coefficient when the random effect contribution and the other design terms are held fixed. Equivalently, it multiplies
\begin{equation}
\frac{r_{ij,\tau}}{1 - r_{ij,\tau}}
= \frac{\xi_{ij,\tau} + 1}{K_{ij} - \xi_{ij,\tau}}
\label{eq:anchor-odds}
\end{equation}
by the exponential of the coefficient. The ratio in \eqref{eq:anchor-odds} describes the anchor's position within its feasible interval; it is not the odds of a binary response event. When a covariate also has a random slope or enters an interaction, its participant-specific effect includes those additional terms.

The transformation in \eqref{eq:conditional-quantile} gives smooth effects on the anchor but stepwise effects on the integer quantile. Different covariate profiles can consequently have different anchors and the same integer quantile. Shape covariates alter the remaining features of the conditional distribution without changing the target anchor when $\eta_{ij,\tau}$ is held fixed. With observation-specific upper bounds, $r_{ij,\tau}$ describes relative position on the normalized scale; comparisons on the original outcome scale must also account for $K_{ij}$.

\subsection{Prior specification and posterior distribution}
\label{sec:priors}

We use proper Gaussian priors for the regression coefficients,
\begin{equation}
\bm{\beta}_\tau \sim \Normal_p\left(\bm{m}_\beta, \bm{V}_\beta\right),
\qquad
\bm{\gamma}_\tau \sim \Normal_d\left(\bm{m}_\gamma, \bm{V}_\gamma\right),
\label{eq:regression-priors}
\end{equation}
where $\bm{m}_\beta$ and $\bm{m}_\gamma$ are fixed prior mean vectors, and $\bm{V}_\beta$ and $\bm{V}_\gamma$ are fixed positive definite prior covariance matrices. Zero means and diagonal covariance matrices give independent centered priors. These hyperparameters are specified on the scales of the predictors used in the model.

For the random effects, separate the standard deviations (SDs) from the correlations:
\begin{equation}
\bm{\Sigma}_\tau
= \bm{D}_\tau\bm{R}_\tau\bm{D}_\tau,
\qquad
\bm{D}_\tau
= \diag\left(\sigma_{1,\tau}, \ldots, \sigma_{q,\tau}\right),
\label{eq:covariance-separation}
\end{equation}
where $\sigma_{\ell,\tau}$ is the SD of the $\ell$th participant effect and $\bm{R}_\tau$ is their correlation matrix. We assign independent priors
\begin{equation}
\sigma_{\ell,\tau} \sim \HalfNormal\left(s_\ell\right),
\quad \ell = 1, \ldots, q,
\qquad
\bm{R}_\tau \sim \LKJ_q\left(\nu\right).
\label{eq:covariance-priors}
\end{equation}
Here, $\HalfNormal\left(s_\ell\right)$ is the distribution of the absolute value of a $\Normal\left(0, s_\ell^2\right)$ variable, with fixed scale hyperparameter $s_\ell > 0$. The Lewandowski, Kurowicka, and Joe (LKJ) prior has fixed concentration hyperparameter $\nu > 0$ and density proportional to $\det\left(\bm{R}_\tau\right)^{\nu - 1}$ over positive definite correlation matrices \citep{lewandowski2009generating}. All prior blocks in \eqref{eq:regression-priors} and \eqref{eq:covariance-priors} are mutually independent. Numerical hyperparameter choices for the application are given in \mainsecref{sec:application}.

When $q = 2$, write $\rho_\tau = \left(\bm{R}_\tau\right)_{12}$ for the correlation between the two participant effects. It has density proportional to $\left(1 - \rho_\tau^2\right)^{\nu - 1}$ on $\left(-1, 1\right)$, or equivalently $\left(\rho_\tau + 1\right)/2 \sim \operatorname{Beta}\left(\nu, \nu\right)$. When $q = 1$, the correlation matrix is fixed at one. A Cholesky factorization $\bm{R}_\tau = \bm{L}_\tau\bm{L}_\tau^{\mathsf T}$, with $\bm{L}_\tau$ lower triangular and having positive diagonal entries, yields the equivalent representation
\begin{equation}
\bm{b}_{i,\tau}
= \bm{D}_\tau\bm{L}_\tau\bm{u}_{i,\tau},
\qquad
\bm{u}_{i,\tau} \sim \Normal_q\left(\bm{0}, \bm{I}_q\right).
\label{eq:noncentered}
\end{equation}
The vectors $\bm{u}_{i,\tau}$ are independent standardized, or whitened, participant effects. The factorization in \eqref{eq:noncentered} ensures a valid covariance representation; the probability model for that covariance remains the prior in \eqref{eq:covariance-priors}.

For the posterior formulation, let $\obs_i \subseteq \left\{1, \ldots, m_i\right\}$ index participant $i$'s observed responses and let $\bm{y}_{\obs}$ collect their observed values $y_{ij}$. For collections of participant vectors, we use $\bm{b}_{1:n,\tau} = \left(\bm{b}_{1,\tau},\ldots,\bm{b}_{n,\tau}\right)$, with analogous notation for other participant-level vectors. Write $\bm{\theta}_\tau = \left(\bm{\beta}_\tau, \bm{\gamma}_\tau, \bm{\sigma}_\tau, \bm{R}_\tau\right)$, with $\bm{\sigma}_\tau = \left(\sigma_{1,\tau}, \ldots, \sigma_{q,\tau}\right)^{\mathsf T}$, and abbreviate the conditional mass function as
\begin{equation}
p_{ij,\tau}\left(y \middle| \bm{b}, \bm{\theta}_\tau\right)
= p_{K_{ij}}\left(y \middle| a_{ij,\tau}, b_{ij,\tau}\right),
\end{equation}
where the shapes on the right are evaluated using $\bm{b}_{i,\tau}=\bm{b}$. For incomplete responses, we assume missing at random (MAR) given the recorded covariates and observed outcomes, with distinct parameters and independent priors for the outcome and missingness mechanisms \citep{rubin1976inference}.

Assuming independence across participants, the joint posterior based on the observed responses is
\begin{equation}
\pi_\tau\left(\bm{\theta}_\tau, \bm{b}_{1,\tau}, \ldots, \bm{b}_{n,\tau}
\middle| \bm{y}_{\obs}\right)
\propto
\pi_\tau\left(\bm{\theta}_\tau\right)
\prod_{i=1}^{n}
\left\{
\varphi_q\left(\bm{b}_{i,\tau}; \bm{0}, \bm{\Sigma}_\tau\right)
\prod_{j \in \obs_i}
p_{ij,\tau}\left(y_{ij} \middle| \bm{b}_{i,\tau}, \bm{\theta}_\tau\right)
\right\}.
\label{eq:posterior}
\end{equation}
Here, $\pi_\tau\left(\bm{\theta}_\tau\right)$ is the joint parameter prior density defined by \eqref{eq:regression-priors} and \eqref{eq:covariance-priors}, and conditioning on $\bm{y}_{\obs}$ denotes the posterior density. With the proper priors above, the posterior is proper for any finite collection of supported observations; the bounded discrete likelihood gives the short argument in \appref{app:propriety}.

Posterior inference for anchors, integer quantiles and other functionals is obtained by evaluating the corresponding expressions at every retained draw after warmup. Parameter summaries use posterior medians and 95\% equal-tailed credible intervals.

\subsection{Quantile inference and comparisons}
\label{sec:multiple-quantiles}

The target in \eqref{eq:conditional-quantile} is conditional on participant effects. At fixed model parameters, marginal quantiles are obtained by averaging the conditional CDF over the distribution of participant effects and then inverting the resulting CDF. Setting the effects to zero gives a conditional reference profile, while averaging conditional quantiles generally does not yield marginal quantiles.

Posterior credible intervals for quantiles describe uncertainty in those quantities. Posterior predictive intervals instead describe variation in replicated outcomes or summaries of those outcomes, incorporating both posterior uncertainty and outcome variability.

A fit anchored at $\tau$ specifies an entire conditional distribution and supplies other quantiles through \eqref{eq:anchor-general}. However, changing the target level generally does not preserve the assumed regression structure. We therefore fit the model separately at prespecified levels $\mathcal{T} \subset \left(0, 1\right)$, using the full observed-data likelihood at each level, with separate $\bm{\beta}_\tau$, $\bm{\gamma}_\tau$, and $\bm{\Sigma}_\tau$.

The models fitted separately at different quantile levels need not represent one common conditional distribution or yield noncrossing anchors. We assess crossing of posterior median anchors and integer quantile summaries at matched covariate profiles, allowing integer ties. Coefficient comparisons across levels are descriptive.

\section{Posterior computation}
\label{sec:computation}

We combine an exact Gaussian coefficient update with elliptical slice sampling \citep{murray2010elliptical}, scalar slice sampling \citep{neal2003slice}, and covariance interweaving \citep{yu2011interweaving}. The coefficient update exploits a relationship between the fixed and random effect designs. We derive it for general compatible designs, then describe the implemented random intercept and slope sampler. Throughout this section, $\tau$ is fixed, and its subscript is suppressed, including in density notation $\pi$.

\subsection{Total participant coefficients}
\label{sec:total-coefficients}

Let $\bm{X}_i \in \R^{m_i \times p}$ and $\bm{Z}_i \in \R^{m_i \times q}$ have rows $\bm{x}_{ij}^{\mathsf T}$ and $\bm{z}_{ij}^{\mathsf T}$, respectively. Suppose that a known $q \times p$ matrix $\bm{H}_i$ satisfies
\begin{equation}
\bm{X}_i = \bm{Z}_i\bm{H}_i.
\label{eq:design-compatibility}
\end{equation}
Using hierarchical centering \citep{gelfand1995efficient}, define the total participant coefficients
\begin{equation}
\bm{c}_i = \bm{H}_i\bm{\beta} + \bm{b}_i,
\qquad
\bm{c}_i \mid \bm{\beta}, \bm{\Sigma}
\sim \Normal_q\left(\bm{H}_i\bm{\beta}, \bm{\Sigma}\right).
\label{eq:total-coefficients}
\end{equation}
This is an invertible change of coordinates with unit Jacobian. Since $\eta_{ij} = \bm{z}_{ij}^{\mathsf T}\bm{c}_i$, participant $i$'s observed log likelihood can be written as
\begin{equation}
\ell_i\left(\bm{c}_i, \bm{\gamma}\right)
= \sum_{j \in \obs_i}
\log p_{K_{ij}}\left(y_{ij} \middle|
a_{ij}\left(\bm{c}_i,\bm{\gamma}\right), b_{ij}\left(\bm{\gamma}\right)\right).
\label{eq:subject-log-likelihood}
\end{equation}
Here, the shape functions are
\begin{equation}
\begin{aligned}
a_{ij}\left(\bm{c}_i, \bm{\gamma}\right)
&= -\frac{\exp\left(\bm{w}_{ij}^{\mathsf T}\bm{\gamma}\right)}
{\log\left\{\expit\left(\bm{z}_{ij}^{\mathsf T}\bm{c}_i\right)\right\}},\\
b_{ij}\left(\bm{\gamma}\right)
&= \frac{\log\left(1 - \tau\right)}
{\log\left[1 - \exp\left\{-\exp\left(\bm{w}_{ij}^{\mathsf T}\bm{\gamma}\right)\right\}\right]}.
\end{aligned}
\label{eq:total-coefficient-shapes}
\end{equation}
These are the shapes in \eqref{eq:conditional-shapes}, expressed in total-coefficient coordinates with $\tau$ fixed. In these coordinates, the observation likelihood does not depend separately on $\bm{\beta}$.

\begin{proposition}[Gaussian coefficient conditional]
\label{prop:gaussian-update}
Under \eqref{eq:design-compatibility} and the Gaussian prior in \eqref{eq:regression-priors}, holding $\bm{c}_{1:n}$ fixed gives
\begin{equation}
\begin{aligned}
\bm{\beta} \mid \bm{c}_{1:n}, \bm{\Sigma}, \bm{\gamma}, \bm{y}_{\obs}
&\sim \Normal_p\left(\bm{Q}_\beta^{-1}\bm{h}_\beta,\bm{Q}_\beta^{-1}\right),\\
\bm{Q}_\beta
&= \bm{V}_\beta^{-1}
 + \sum_{i=1}^{n}\bm{H}_i^{\mathsf T}\bm{\Sigma}^{-1}\bm{H}_i,\\
\bm{h}_\beta
&= \bm{V}_\beta^{-1}\bm{m}_\beta
 + \sum_{i=1}^{n}\bm{H}_i^{\mathsf T}\bm{\Sigma}^{-1}\bm{c}_i.
\end{aligned}
\label{eq:beta-gibbs}
\end{equation}
The precision $\bm{Q}_\beta$ is positive definite whenever $\bm{V}_\beta$ and $\bm{\Sigma}$ are positive definite.
\end{proposition}
The derivation is given in \appref{app:gaussian-derivation}. The update changes $\bm{b}_i = \bm{c}_i - \bm{H}_i\bm{\beta}$ while preserving every conditional predictor. The Gaussian conditional follows from the hierarchical normal distribution of the total coefficients; it does not approximate the discrete observation likelihood. Numerically, the draw is obtained by a Cholesky factorization of $\bm{Q}_\beta$ and triangular solves, without forming its inverse.

Condition \eqref{eq:design-compatibility} holds for a fixed time slope and participant-level covariates when the random effects include an intercept and the same time slope. It need not hold for arbitrary designs in \mainsecref{sec:general-specification}. Outside this class, \eqref{eq:beta-gibbs} is unavailable for the full coefficient vector, and a likelihood-dependent update, such as elliptical slice sampling under the Gaussian prior, is required. Positive definiteness of the posterior precision under a proper prior does not itself establish likelihood identifiability.

\subsection{Participant and shape updates}
\label{sec:slice-updates}

At fixed $\bm{\beta}$, $\bm{\Sigma}$, and $\bm{\gamma}$, the participant effects have conditionally independent densities
\begin{equation}
\pi\left(\bm{b}_i \middle| \bm{\beta},\bm{\Sigma},\bm{\gamma},\bm{y}_{\obs}\right)
\propto \exp\left\{\ell_i\left(\bm{H}_i\bm{\beta}+\bm{b}_i,\bm{\gamma}\right)\right\}
\varphi_q\left(\bm{b}_i;\bm{0},\bm{\Sigma}\right),
\label{eq:effect-conditional}
\end{equation}
where $\ell_i$ is participant $i$'s observed log likelihood defined in \eqref{eq:subject-log-likelihood}. An independent auxiliary draw $\bm{v}_i \sim \Normal_q\left(\bm{0},\bm{\Sigma}\right)$ defines elliptical proposals indexed by an angle $\omega$,
\begin{equation}
\bm{b}_i\left(\omega\right)
= \bm{b}_i\cos\left(\omega\right)+\bm{v}_i\sin\left(\omega\right).
\label{eq:effect-ellipse}
\end{equation}
For an independent $U_i\sim\Uniform\left(0,1\right)$, let
\begin{equation}
h_i=\ell_i\left(\bm{H}_i\bm{\beta}+\bm{b}_i,\bm{\gamma}\right)+\log U_i.
\label{eq:effect-slice-height}
\end{equation}
Draw an initial angle $\omega_0 \sim \Uniform\left(0, 2\pi\right)$ and initialize the angular bracket to $\left[\omega_0 - 2\pi, \omega_0\right]$. Starting with $\omega = \omega_0$, accept the proposal when $\ell_i\left(\bm{H}_i\bm{\beta} + \bm{b}_i\left(\omega\right), \bm{\gamma}\right) > h_i$. Otherwise, replace the bracket endpoint having the same sign as $\omega$ by $\omega$, draw a new angle uniformly from the reduced bracket, and repeat. The Gaussian prior is accounted for by the elliptical construction, so the slice comparison uses only the observed log likelihood \citep{murray2010elliptical}. After accepting the move, set $\bm{c}_i = \bm{H}_i\bm{\beta} + \bm{b}_i$.

Holding $\bm{c}_{1:n}$ fixed, the shape coefficients have conditional density
\begin{equation}
\pi\left(\bm{\gamma}\middle|\bm{c}_{1:n},\bm{\beta},\bm{\Sigma},\bm{y}_{\obs}\right)
\propto \exp\left\{\sum_{i=1}^{n}\ell_i\left(\bm{c}_i,\bm{\gamma}\right)\right\}
\varphi_d\left(\bm{\gamma};\bm{m}_\gamma,\bm{V}_\gamma\right).
\label{eq:shape-conditional}
\end{equation}
We update this block jointly using the prior-centered ellipse
\begin{equation}
\bm{\gamma}\left(\omega\right)
=\bm{m}_\gamma+\left(\bm{\gamma}-\bm{m}_\gamma\right)\cos\left(\omega\right)
+\bm{v}_\gamma\sin\left(\omega\right),
\qquad \bm{v}_\gamma\sim\Normal_d\left(\bm{0},\bm{V}_\gamma\right),
\label{eq:shape-ellipse}
\end{equation}
with a slice height based on $\sum_i\ell_i$. The implementation uses the centered prior case $\bm{m}_\gamma=\bm{0}$.

\subsection{Covariance updates in two parameterizations}
\label{sec:covariance-updates}

For the implemented case $q = 2$, use unconstrained coordinates
\begin{equation}
\bm{\psi}=\left(\lambda_1,\lambda_2,\zeta\right)^{\mathsf T}
=\left(\log\sigma_1,\log\sigma_2,\operatorname{atanh}\rho\right)^{\mathsf T}.
\label{eq:covariance-coordinates}
\end{equation}
Here, $\operatorname{atanh}$ is the inverse hyperbolic tangent. Write $\bm{\Sigma}\left(\bm{\psi}\right)$ for the covariance in \eqref{eq:covariance-separation} recovered using $\sigma_\ell=\exp\left(\lambda_\ell\right)$ and $\rho=\tanh\left(\zeta\right)$. The log prior density with respect to $\mathrm{d}\bm{\psi}$ is, up to a constant,
\begin{equation}
g\left(\bm{\psi}\right)
=\lambda_1+\lambda_2
-\frac{1}{2}\sum_{\ell=1}^{2}\frac{\exp\left(2\lambda_\ell\right)}{s_\ell^2}
+\nu\log\left(1-\tanh^2\zeta\right).
\label{eq:transformed-covariance-prior}
\end{equation}
The transformation Jacobian is $\sigma_1\sigma_2\left(1-\rho^2\right)$, giving exponent $\nu$, rather than $\nu-1$, in the final term. The derivation and the explicit covariance Cholesky factor are given in \appref{app:covariance}.

Holding $\bm{b}_{1:n}$ and $\bm{\beta}$ fixed, successive scalar slice updates of the three coordinates target the centered log conditional
\begin{equation}
t_{\mathrm{C}}\left(\bm{\psi}\right)
=g\left(\bm{\psi}\right)
-\frac{n}{2}\log\det\bm{\Sigma}\left(\bm{\psi}\right)
-\frac{1}{2}\operatorname{tr}\left\{
\bm{\Sigma}\left(\bm{\psi}\right)^{-1}\bm{S}_b\right\},
\quad \bm{S}_b=\sum_{i=1}^{n}\bm{b}_i\bm{b}_i^{\mathsf T}.
\label{eq:centered-covariance-target}
\end{equation}
Here, $\operatorname{tr}$ denotes the matrix trace. The observation likelihood is constant during this update because the linear predictors are fixed.

Next, write $\bm{\Sigma}=\bm{C}\bm{C}^{\mathsf T}$ with $\bm{C}=\bm{C}\left(\bm{\psi}\right)=\bm{D}\bm{L}$, the lower triangular Cholesky factor of the full covariance, and set $\bm{u}_i=\bm{C}^{-1}\bm{b}_i$. Holding these whitened effects, $\bm{\beta}$, and $\bm{\gamma}$ fixed, a second sequence of scalar slice updates targets
\begin{equation}
t_{\mathrm{N}}\left(\bm{\psi}\right)
=g\left(\bm{\psi}\right)
+\sum_{i=1}^{n}\ell_i\left(
\bm{H}_i\bm{\beta}+\bm{C}\left(\bm{\psi}\right)\bm{u}_i,
\bm{\gamma}\right).
\label{eq:noncentered-covariance-target}
\end{equation}
For each participant, the transformation $\bm{b}_i = \bm{C}\bm{u}_i$ has Jacobian $\det\left(\bm{C}\right) = \det\left(\bm{\Sigma}\right)^{1/2}$, which cancels the factor $\det\left(\bm{\Sigma}\right)^{-1/2}$ in the Gaussian density. The resulting standard normal density of $\bm{u}_i$ is independent of $\bm{\psi}$. Consequently, no additional determinant term appears in \eqref{eq:noncentered-covariance-target}, and the likelihood is recomputed as $\bm{\psi}$ changes. Transforming back moves the covariance and participant effects together. These complementary representations implement the interweaving principle of Yu and Meng~\citep{yu2011interweaving}.

\floatstyle{ruled}
\restylefloat{algorithm}
\captionsetup[algorithm]{labelsep=period}

\begin{algorithm}[tbp]
\caption{One posterior sampling sweep}
\label{alg:sampler}
\begin{minipage}{\linewidth}
\small
For a fixed quantile level $\tau$, take the observed data, prior hyperparameters, and a valid current state $\left(\bm{c}_{1:n}, \bm{\beta}, \bm{\gamma}, \bm{\psi}\right)$ for the random intercept and slope model, with $\bm{X}_i = \bm{Z}_i\bm{H}_i$.

\begin{enumerate}[leftmargin=*, label=\arabic*., itemsep=5pt, topsep=5pt]
\item For each participant, update $\bm{b}_i = \bm{c}_i - \bm{H}_i\bm{\beta}$ by elliptical slice sampling using \eqref{eq:effect-ellipse} and \eqref{eq:effect-slice-height}. Set $\bm{c}_i = \bm{H}_i\bm{\beta} + \bm{b}_i$.

\item Hold $\bm{c}_{1:n}$ fixed and draw $\bm{\beta}$ from \eqref{eq:beta-gibbs}. Recompute $\bm{b}_i = \bm{c}_i - \bm{H}_i\bm{\beta}$ for every participant.

\item Hold $\bm{b}_{1:n}$ and $\bm{\beta}$ fixed and update each coordinate of $\bm{\psi}$ by scalar slice sampling with the centered log target in \eqref{eq:centered-covariance-target}.

\item Compute $\bm{u}_i = \bm{C}\left(\bm{\psi}\right)^{-1}\bm{b}_i$ for every participant. Hold these whitened effects, $\bm{\beta}$, and $\bm{\gamma}$ fixed and update each coordinate of $\bm{\psi}$ by scalar slice sampling with the log target in \eqref{eq:noncentered-covariance-target}.

\item For every participant, set $\bm{b}_i = \bm{C}\left(\bm{\psi}\right)\bm{u}_i$ and $\bm{c}_i = \bm{H}_i\bm{\beta} + \bm{b}_i$.

\item Hold $\bm{c}_{1:n}$ fixed and jointly update $\bm{\gamma}$ using \eqref{eq:shape-ellipse} and a slice based on $\sum_{i=1}^{n}\ell_i\left(\bm{c}_i, \bm{\gamma}\right)$.
\end{enumerate}

Repeat the sweep and retain every iteration after warmup.
\end{minipage}
\end{algorithm}

Each stochastic step in \mainalgref{alg:sampler} preserves its conditional target, and each coordinate change is invertible with the stated Jacobian. Their composition therefore leaves \eqref{eq:posterior} invariant; \appref{app:invariance} gives the argument. This establishes the target of the sampling procedure, while convergence and Monte Carlo precision for a finite run require separate assessment.

\subsection{Implementation and numerical assessment}

The sampling sweeps are implemented in C++, with R coordinating independent chains, posterior summaries, and predictive checks. The model specification and settings supported by the supplied application code are described in \mainsecref{sec:application}.

Probability calculations use logarithms of differences of survival probabilities, explicit endpoint cases, and stable evaluations of logarithms near zero. Limiting expressions are used where direct evaluation would lose floating-point precision. Scalar slice updates use randomized starting brackets, stepping out and shrinkage, with fixed initial widths. Numerical iteration limits raise an error if reached. Independent chains can run in parallel; checkpoints retain the sampler state and random-number generator state for restarting. All post-warmup draws contribute to posterior summaries and prediction.

Implementation checks compare the probability calculations with a separate R implementation, verify the covariance density transformations, and assess sampler moments in targets with analytically known moments. Sampling is assessed using trace plots, the rank-normalized convergence diagnostic $\widehat{R}$, and bulk and tail effective sample sizes \citep{vehtari2021rank}. Parameter recovery and interval coverage are evaluated in \mainsecref{sec:simulation}.

Posterior predictive checks compare replicated and observed score frequencies \citep{gelman1996posterior}. Conditional replicates retain the sampled effects of existing participants. Replicates for a new cohort instead draw one fresh effect vector per participant at each posterior draw and share it across that participant's assessments. Both schemes retain the observed covariates, assessment times, and pattern of recorded outcomes. These checks do not test the MAR assumption or constitute out-of-sample validation.

\section{Simulation study}
\label{sec:simulation}

\subsection{Design and data generation}

We evaluated parameter recovery and the repeated-sampling coverage of Bayesian credible intervals under the proposed model. Four scenarios crossed $n \in \left\{100, 200\right\}$ participants with target quantile levels $\tau \in \left\{0.50, 0.90\right\}$. We generated 300 independent datasets per scenario. The generating model used the Benton Visual Retention Test (BVRT) specification in \eqref{eq:bvrt-design}, with scores on $\left\{0, \ldots, 15\right\}$, quantile regression terms for time, education, entry age, and sex, a correlated Gaussian random intercept and time slope, and an education effect in the shape regression. For each quantile level, generating values were the corresponding application posterior medians rounded to two decimal places and held fixed across datasets and sample sizes (\maintabref{tab:bvrt-simulation}). The two quantile settings therefore represent distinct application-based distributions, rather than two targets within a single generating distribution.

Covariates and assessment schedules were sampled jointly from the 500 participants in the PAQUID subsample described in \mainsecref{sec:application}. For each simulated dataset, we sampled $n$ participants uniformly with replacement, retaining each selected participant's education, sex, entry age, and assessment times at which a BVRT score was observed. Elapsed time was expressed per five years, and entry age was centered at the mean of the original 500 participants and expressed per ten years. This centering constant was used for both generation and fitting and was not recomputed within simulated samples. Sampling whole participant records preserved the empirical relationships between covariates, follow-up times, and numbers of observed assessments, including participants with only one assessment.

For every selected participant, we independently generated a new Gaussian effect vector using the scenario-specific SDs and correlation. Scores were then generated independently conditional on this vector and the covariates, using the inverse CDF of the discrete Kumaraswamy model. Repeated selections of the same source participant received independent effects and scores. The resampling step retained assessment times with observed BVRT scores; new responses were generated from the specified model. Thus, the study conditioned on sampled observation patterns and did not generate a dropout or missingness mechanism.

\subsection{Estimation and results}

Each dataset was fitted at its generating quantile level using the same likelihood and prior specification as the application: independent zero-mean Gaussian priors with SD 10 for the quantile and shape coefficients, half-normal priors with scale 5 for both participant-effect SDs, and a uniform prior on their correlation. Two chains with different initial states each used 1,000 warmup iterations followed by 10,000 retained iterations. The first 100 warmup iterations used only the centered covariance update; covariance interweaving was enabled for the remaining warmup and all retained iterations. Initial states did not use the generating parameters or simulated participant effects. Posterior medians and 95\% equal-tailed credible intervals used all 20,000 post-warmup draws.

Recovery was assessed for the five components of $\bm{\beta}_\tau$, the two components of $\bm{\gamma}_\tau$, the two SDs $\sigma_{1,\tau}$ and $\sigma_{2,\tau}$, and the correlation $\rho_\tau$. Bias was the average posterior median minus the generating value. Root mean squared error (RMSE) was the square root of the average squared difference between the posterior median and the generating value. Coverage probability (CP) was the percentage of 95\% credible intervals containing the generating value. All three measures were calculated separately for each parameter and scenario, with bias and RMSE expressed on the parameter's original scale.

All 300 datasets were fitted in each scenario, giving 1,200 completed fits. All completed fits contributed to \maintabref{tab:bvrt-simulation}, with no exclusions.

Quantile and shape regression coefficients showed small biases and generally near-nominal coverage of 95\% credible intervals. Random slope SDs tended to be underestimated, while negative correlations were estimated closer to zero. Correlation bias decreased with sample size, although coverage remained above the nominal level. RMSE decreased with sample size for every parameter at both quantile levels.

\afterpage{\begin{landscape}
\begingroup
\fontsize{10}{12}\selectfont
\setlength{\tabcolsep}{5pt}
\renewcommand{\arraystretch}{1.0}
\setlength{\LTleft}{\fill}
\setlength{\LTright}{\fill}
\begin{longtable}{@{}>{\arraybackslash}m{1.1cm}>{\arraybackslash}m{2.1cm}>{\arraybackslash}m{4.2cm}*{7}{>{\raggedleft\arraybackslash}m{1.6cm}}@{}}
\caption{Bias, RMSE and coverage for the BVRT quantile mixed models.}\label{tab:bvrt-simulation}\\
\toprule
 & & & & \multicolumn{3}{c}{$\bm{n = 100}$} & \multicolumn{3}{c}{$\bm{n = 200}$} \\
\cmidrule(lr){5-7}\cmidrule(lr){8-10}
$\bm{\tau}$ & \textbf{Parameter} & \textbf{Description} & \textbf{True} & \textbf{Bias} & \textbf{RMSE} & \textbf{CP (\%)} & \textbf{Bias} & \textbf{RMSE} & \textbf{CP (\%)} \\
\midrule
\endfirsthead
\multicolumn{10}{c}{\tablename\ \thetable{} (continued)} \\
\toprule
 & & & & \multicolumn{3}{c}{$\bm{n = 100}$} & \multicolumn{3}{c}{$\bm{n = 200}$} \\
\cmidrule(lr){5-7}\cmidrule(lr){8-10}
$\bm{\tau}$ & \textbf{Parameter} & \textbf{Description} & \textbf{True} & \textbf{Bias} & \textbf{RMSE} & \textbf{CP (\%)} & \textbf{Bias} & \textbf{RMSE} & \textbf{CP (\%)} \\
\midrule
\endhead
\midrule
\multicolumn{10}{r}{\textit{Continued on next page}} \\
\endfoot
\bottomrule
\multicolumn{10}{@{}p{\dimexpr18.6cm+18\tabcolsep\relax}@{}}{\fontsize{9}{11}\selectfont \textit{Notes:} True values are application posterior medians rounded to two decimal places. Bias is the average posterior median minus truth. RMSE is the square root of the average squared difference between the posterior median and truth. CP is coverage (\%) of 95\% equal-tailed credible intervals. All 20,000 post-warmup draws per fit are used. Each scenario includes all 300 completed fits, including those with diagnostic flags. No fits were excluded. Parameters use the application scales.} \\
\endlastfoot
\multicolumn{10}{@{}l}{\textit{Quantile regression}} \\*
\addlinespace[0.1em]
$0.50$ & $\beta_{0,\tau}$ & Intercept & $0.60$ & $0.014$ & $0.135$ & $94.0$ & $-0.001$ & $0.091$ & $96.0$ \\
 & $\beta_{t,\tau}$ & Time (per 5 years) & $-0.17$ & $-0.003$ & $0.034$ & $97.3$ & $-0.001$ & $0.024$ & $94.3$ \\
 & $\beta_{E,\tau}$ & Primary school graduation & $0.58$ & $-0.021$ & $0.145$ & $96.7$ & $-0.003$ & $0.101$ & $95.0$ \\
 & $\beta_{A,\tau}$ & Entry age (per 10 years) & $-0.46$ & $-0.004$ & $0.105$ & $95.7$ & $0.001$ & $0.080$ & $92.7$ \\
 & $\beta_{M,\tau}$ & Male sex & $0.14$ & $0.010$ & $0.132$ & $95.3$ & $0.009$ & $0.087$ & $95.7$ \\
\addlinespace[0.25em]
\multicolumn{10}{@{}l}{\textit{Shape regression}} \\*
\addlinespace[0.1em]
$0.50$ & $\gamma_{0,\tau}$ & Intercept & $0.92$ & $-0.012$ & $0.102$ & $94.7$ & $-0.010$ & $0.078$ & $93.0$ \\
 & $\gamma_{E,\tau}$ & Primary school graduation & $-0.13$ & $0.009$ & $0.119$ & $95.3$ & $0.001$ & $0.086$ & $94.3$ \\
\addlinespace[0.25em]
\multicolumn{10}{@{}l}{\textit{Participant effects}} \\*
\addlinespace[0.1em]
$0.50$ & $\sigma_{1,\tau}$ & Random intercept SD & $0.56$ & $-0.019$ & $0.073$ & $95.3$ & $-0.013$ & $0.052$ & $96.7$ \\
 & $\sigma_{2,\tau}$ & Random slope SD & $0.14$ & $-0.021$ & $0.056$ & $96.7$ & $-0.014$ & $0.045$ & $92.0$ \\
 & $\rho_\tau$ & Intercept-slope correlation & $-0.54$ & $0.123$ & $0.297$ & $98.3$ & $0.055$ & $0.208$ & $97.0$ \\
\addlinespace[0.4em]
\multicolumn{10}{@{}l}{\textit{Quantile regression}} \\*
\addlinespace[0.1em]
$0.90$ & $\beta_{0,\tau}$ & Intercept & $1.29$ & $0.018$ & $0.145$ & $95.0$ & $0.004$ & $0.101$ & $95.7$ \\
 & $\beta_{t,\tau}$ & Time (per 5 years) & $-0.16$ & $-0.003$ & $0.034$ & $96.3$ & $0.000$ & $0.025$ & $93.7$ \\
 & $\beta_{E,\tau}$ & Primary school graduation & $0.63$ & $0.001$ & $0.160$ & $93.7$ & $0.001$ & $0.116$ & $94.3$ \\
 & $\beta_{A,\tau}$ & Entry age (per 10 years) & $-0.42$ & $-0.001$ & $0.100$ & $96.0$ & $-0.002$ & $0.071$ & $93.7$ \\
 & $\beta_{M,\tau}$ & Male sex & $0.13$ & $-0.005$ & $0.120$ & $95.0$ & $0.002$ & $0.086$ & $95.0$ \\
\addlinespace[0.25em]
\multicolumn{10}{@{}l}{\textit{Shape regression}} \\*
\addlinespace[0.1em]
$0.90$ & $\gamma_{0,\tau}$ & Intercept & $0.35$ & $-0.028$ & $0.179$ & $93.3$ & $-0.007$ & $0.133$ & $92.3$ \\
 & $\gamma_{E,\tau}$ & Primary school graduation & $-0.23$ & $0.020$ & $0.198$ & $95.7$ & $-0.001$ & $0.152$ & $92.0$ \\
\addlinespace[0.25em]
\multicolumn{10}{@{}l}{\textit{Participant effects}} \\*
\addlinespace[0.1em]
$0.90$ & $\sigma_{1,\tau}$ & Random intercept SD & $0.52$ & $-0.019$ & $0.068$ & $96.0$ & $-0.013$ & $0.052$ & $95.0$ \\
 & $\sigma_{2,\tau}$ & Random slope SD & $0.13$ & $-0.020$ & $0.052$ & $96.0$ & $-0.017$ & $0.044$ & $93.0$ \\
 & $\rho_\tau$ & Intercept-slope correlation & $-0.55$ & $0.128$ & $0.300$ & $97.3$ & $0.077$ & $0.226$ & $97.0$ \\
\end{longtable}
\endgroup
\end{landscape}}

\section{Application to longitudinal cognitive scores}
\label{sec:application}

\subsection{Data and model specification}

We illustrate the model using repeated BVRT scores from PAQUID, a prospective observational study of aging in southwestern France \citep{letenneur1994incidence}. The analysis uses the 500-participant subsample distributed as \texttt{paquid} in the \texttt{lcmm} R package \citep{proustlima2017estimation}. BVRT measures visual memory on an integer scale from 0 to 15, with higher scores indicating better performance. Of the 2,250 supplied assessment records, 1,950 had an observed BVRT score. All 500 participants contributed to the analysis, including 115 with a single observed score. The sample included 288 women and 355 primary school graduates; mean age at cohort entry was 74.2 years. Assessment times and the number of observations varied between participants. Inference used the observed responses under the ignorable MAR assumption stated in \mainsecref{sec:priors}, without imputing missing scores.

We fitted the model separately at $\tau \in \left\{0.10, 0.25, 0.50, 0.75, 0.90\right\}$, with $K_{ij} = 15$. The design vectors in \eqref{eq:mixed-predictor} were
\begin{equation}
\begin{aligned}
\bm{x}_{ij} &= \left(1, t_{ij}, E_i, A_i, M_i\right)^{\mathsf T},\\
\bm{z}_{ij} &= \left(1, t_{ij}\right)^{\mathsf T},
\qquad
\bm{w}_{ij} = \left(1, E_i\right)^{\mathsf T},
\end{aligned}
\label{eq:bvrt-design}
\end{equation}
where $t_{ij}$ is years since cohort entry divided by five, $E_i$ indicates primary school graduation, $A_i$ is entry age centered at its participant-level sample mean and divided by ten, and $M_i$ indicates male sex. Thus, the quantile predictor included time, education, entry age, and sex, with a correlated random intercept and time slope. The corresponding SDs are $\sigma_{1,\tau}$ for the random intercept and $\sigma_{2,\tau}$ for the time slope, with correlation $\rho_\tau$ as defined in \mainsecref{sec:priors}. Education also entered the shape predictor. Separating entry age from elapsed time distinguishes age differences between participants from longitudinal change within participants.

To identify the coefficients by their covariates, we write
\begin{equation}
\begin{aligned}
\bm{\beta}_\tau
&= \left(\beta_{0,\tau}, \beta_{t,\tau}, \beta_{E,\tau}, \beta_{A,\tau}, \beta_{M,\tau}\right)^{\mathsf T},\\
\bm{\gamma}_\tau
&= \left(\gamma_{0,\tau}, \gamma_{E,\tau}\right)^{\mathsf T}.
\end{aligned}
\label{eq:bvrt-coefficient-notation}
\end{equation}
Here, $\beta_{0,\tau}$ is the quantile intercept, and $\beta_{t,\tau}$, $\beta_{E,\tau}$, $\beta_{A,\tau}$, and $\beta_{M,\tau}$ are the coefficients for elapsed time, primary school graduation, entry age, and male sex, respectively. These coefficients act on the logit of the scaled quantile anchor. The coefficients $\gamma_{0,\tau}$ and $\gamma_{E,\tau}$ are the intercept and education effect in the regression for $\log\phi_{ij,\tau}$.

For the application, the regression prior hyperparameters in \eqref{eq:regression-priors} were
\begin{equation}
\bm{m}_\beta = \bm{0}_5,
\qquad \bm{V}_\beta = 100\bm{I}_5,
\qquad \bm{m}_\gamma = \bm{0}_2,
\qquad \bm{V}_\gamma = 100\bm{I}_2.
\label{eq:bvrt-regression-hyperparameters}
\end{equation}
These choices give all quantile and shape regression coefficients independent zero-mean Gaussian priors with SD 10. In \eqref{eq:covariance-priors}, we set $s_1 = s_2 = 5$ and $\nu = 1$, giving half-normal priors with scale 5 for the random intercept and slope SDs and a uniform prior on their correlation over $\left(-1, 1\right)$. These choices specify diffuse proper priors on the regression coefficients and random-effect SDs, with the same hyperparameters at all five quantile levels. Their implications for the Kumaraswamy shape parameters depend on the nonlinear parameterization and the chosen quantile level $\tau$.

The supplied code implements this application design, with two random effects, five quantile regression coefficients, two shape regression coefficients with a binary predictor, and a common upper bound of 15. The initial scalar slice widths for $\lambda_1$, $\lambda_2$, and $\zeta$ are 0.35, 0.35, and 0.8, respectively.

Using the sampler in \mainsecref{sec:computation}, two chains each retained 20,000 iterations after 2,000 warmup iterations for each fit. Summaries used all 40,000 retained draws. Across all five fits, rank-normalized $\widehat{R}$ values were below 1.01 and bulk and tail effective sample sizes exceeded 900 for all monitored parameters, including the participant effects.

\subsection{Results}

Table~\ref{tab:bvrt-parameters} in \suppsecref{supp:parameter-summaries} of the Supplementary Material reports posterior medians and 95\% equal-tailed credible intervals for the model parameters at each fitted quantile level.

\begin{figure}[!t]
\centering
\includegraphics[width=\textwidth]{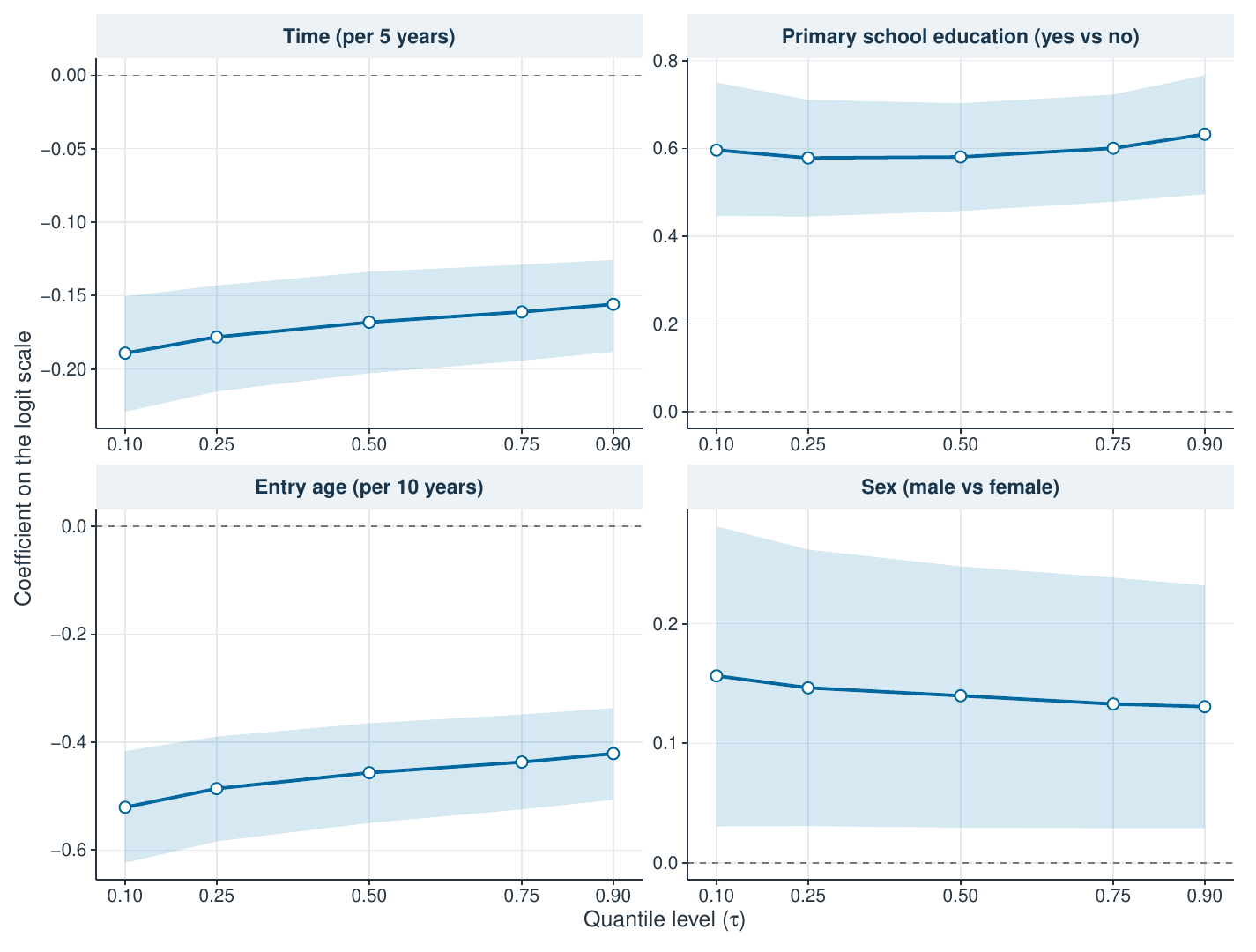}
\caption{Covariate associations with conditional BVRT quantile anchors. Points show posterior medians of the regression coefficients, and shaded regions connect their pointwise 95\% equal-tailed credible intervals at the five fitted quantile levels. Coefficients act on the logit of the scaled anchor. Time is expressed per five years and entry age per ten years; the reference categories are no primary school graduation and female sex. The horizontal dashed lines indicate zero. Connecting lines and shading are visual guides, do not represent estimates at unfitted quantile levels, and do not form a simultaneous credible band.}
\label{fig:bvrt-coefficients}
\end{figure}

\mainfigref{fig:bvrt-coefficients} shows that elapsed time and older entry age were associated with lower conditional quantile anchors, whereas primary school graduation and male sex were associated with higher anchors. Pointwise 95\% credible intervals excluded zero for all four coefficients at every fitted level. On the logit scale, the negative associations with time and entry age were more pronounced at lower quantile levels. The education coefficient was broadly stable, while the male-female contrast decreased slightly towards the upper quantiles. Comparisons across levels are descriptive because the models were fitted separately; these coefficients describe associations with the scaled anchor, not changes measured in BVRT score points.

Across all five fitted quantile levels, the model indicated variation between participants in both intercepts and time slopes, with negative correlations between these effects. Primary school graduation was also associated with a lower shape parameter, indicating differences in distributional shape beyond the association with the quantile anchor.

For each participant, the grid combined observed assessment times with 21 equally spaced times between their first and last observed assessments. Reference grids used 81 equally spaced times from cohort entry to the maximum observed follow-up, at mean entry age, for each education and sex combination, with participant effects set to zero. No crossing was found among the posterior median anchors or integer quantiles at the observed assessments or the evaluated participant and reference-profile time grids. Participant-specific comparisons used the random effects estimated separately at each fitted quantile level. These findings do not guarantee noncrossing at other points or for individual posterior draws.

Figures~\ref{fig:bvrt-frequencies-q10}--\ref{fig:bvrt-frequencies-sex-q90} in \suppsecref{supp:score-frequencies} of the Supplementary Material present posterior predictive score frequencies for all five fitted quantile levels, overall and by education and sex. Both prediction schemes retain the observed covariates and assessment pattern. For the model fitted at $\tau = 0.10$, conditional and new-cohort predictions reproduced the broad score distribution and gave similar predictive medians overall. Local discrepancies remained, particularly among primary school graduates: frequencies at scores 10 and 11 were underpredicted, whereas those at scores 12 and 13 were overpredicted. At all five fitted quantile levels, the predictive median ceiling frequency was below the observed proportion for both prediction schemes, overall and within each education and sex group. These checks indicate that the model captures the broad distributional pattern but does not fully reproduce the upper-score probabilities.

\FloatBarrier

\section{Discussion}
\label{sec:discussion}

The proposed model provides a Bayesian quantile regression framework for bounded integer longitudinal outcomes. Its smooth quantile anchor supports regression modeling while retaining an explicit relationship with the integer quantile. Shape regression and correlated participant effects accommodate distributional differences and dependence within participants. Posterior computation combines established sampling methods with an exact Gaussian coefficient update available under compatible fixed and random effect designs.

Simulations showed small regression coefficient biases and generally near-nominal interval coverage. Covariance parameters were more difficult to estimate, although increasing the sample size reduced RMSE for every parameter. These results concern correctly specified models calibrated to one application; robustness to misspecification and comparative computational efficiency remain to be evaluated. Because generating parameters differed between quantile settings, performance differences cannot be attributed to quantile level alone.

The BVRT application illustrated covariate associations across five conditional quantile levels. Posterior predictive checks captured broad distributional patterns but underpredicted ceiling frequencies. Interpretation remains associational and depends on the stated MAR assumption. Informative attendance and truncation by death require separate consideration. Because the models are fitted separately, their target quantiles are not constrained to arise from a single common conditional distribution. The absence of crossing among the evaluated summaries does not guarantee noncrossing elsewhere.

Future work could consider joint estimation across quantile levels and more flexible distributions near the boundaries. Mixtures of discrete Kumaraswamy components on the same support, whose smooth extensions share a common target anchor, would preserve that anchor and its integer quantile while allowing more flexible probability masses. Identifiability and computation for such extensions require further study.

\FloatBarrier

\section*{Author contributions}

\ifAnonymous
Statement withheld for anonymous review.
\else
The sole author was responsible for the conceptualization, methodology, software development, formal analysis, and preparation and revision of the manuscript.
\fi

\section*{Disclosure statement}

\ifAnonymous
Statement withheld for anonymous review.
\else
The author declares no competing interests.
\fi

\section*{Data and code availability}

The data analyzed in this study are publicly available as the \texttt{paquid} dataset distributed with the \texttt{lcmm} R package \citep{proustlima2017estimation}, available from \url{https://CRAN.R-project.org/package=lcmm}. This dataset is a subsample of the PAQUID cohort described by Letenneur et al.~\citep{letenneur1994incidence}; the full cohort is not distributed with the package.

\ifAnonymous
Code repository details are withheld for anonymous review.
\else
The R and C++ code for the application analysis, simulation study, and manuscript outputs is publicly available at \url{https://github.com/DABURGER1/Discrete_Bounded_QRMM}.
\fi

\bibliographystyle{tfnlm}
\bibliography{bib}

\FloatBarrier
\appendix
\titleformat{\section}{\normalfont\large\bfseries}{Appendix \thesection.}{0.6em}{}
\setcounter{equation}{0}
\renewcommand{\theequation}{\thesection.\arabic{equation}}
\renewcommand{\theHequation}{appendix.\thesection.\arabic{equation}}

\section{Posterior propriety and Gaussian update}
\label{app:gaussian}

\subsection{Posterior propriety}
\label{app:propriety}

For fixed $\tau\in\left(0,1\right)$, let $\mathrm{d}\Pi_0$ denote the joint prior probability measure for $\bm{\theta}_\tau$ and $\bm{b}_{1:n,\tau}$, including the conditional Gaussian distribution of the participant effects. With supported observations and finite parameter values in the prior support, every observed mass is strictly positive and at most one. Hence
\begin{equation}
0 < L_{\mathrm{cond}}
:=\prod_{i=1}^{n}\prod_{j\in\obs_i}
p_{ij,\tau}\left(y_{ij}\middle|\bm{b}_{i,\tau},\bm{\theta}_\tau\right)
\leq 1.
\end{equation}
All prior distributions are proper, so the posterior normalizing constant satisfies
\begin{equation}
0 < Z_\tau:=\int L_{\mathrm{cond}}\,\mathrm{d}\Pi_0\leq\int\mathrm{d}\Pi_0=1.
\label{eq:posterior-normalization}
\end{equation}
Thus \eqref{eq:posterior} defines a proper posterior. Integrating out the participant effects gives the same constant for the marginal likelihood formulation. This argument establishes normalizability, not likelihood identifiability or insensitivity to the prior in a weakly informed model.

\subsection{Derivation of Proposition~\ref*{prop:gaussian-update}}
\label{app:gaussian-derivation}

Suppress $\tau$ as in \mainsecref{sec:computation}. For fixed covariance parameters, the map from $\left(\bm{\beta},\bm{b}_{1:n}\right)$ to $\left(\bm{\beta},\bm{c}_{1:n}\right)$ has block triangular derivative matrix with identity diagonal blocks. Its determinant is one, even though each $\bm{c}_i$ depends on $\bm{\beta}$. Consequently, the posterior in total-coefficient coordinates is proportional to
\begin{equation}
\pi\left(\bm{\beta}\right)\pi\left(\bm{\gamma}\right)
\pi\left(\bm{\sigma},\bm{R}\right)
\prod_{i=1}^{n}
\left[
\exp\left\{\ell_i\left(\bm{c}_i,\bm{\gamma}\right)\right\}
\varphi_q\left(\bm{c}_i;\bm{H}_i\bm{\beta},\bm{\Sigma}\right)
\right].
\label{eq:total-posterior}
\end{equation}
At fixed $\bm{c}_{1:n}$, the likelihood factors in \eqref{eq:total-posterior} are constant in $\bm{\beta}$. The log conditional is therefore
\begin{equation}
\begin{aligned}
\log\pi\left(\bm{\beta}\middle|\bm{c}_{1:n},\bm{\Sigma},\bm{\gamma},\bm{y}_{\obs}\right)
&= -\frac{1}{2}\left(\bm{\beta}-\bm{m}_\beta\right)^{\mathsf T}
\bm{V}_\beta^{-1}\left(\bm{\beta}-\bm{m}_\beta\right)\\
&\quad -\frac{1}{2}\sum_{i=1}^{n}
\left(\bm{c}_i-\bm{H}_i\bm{\beta}\right)^{\mathsf T}
\bm{\Sigma}^{-1}\left(\bm{c}_i-\bm{H}_i\bm{\beta}\right)
+\mathrm{const}\\
&= -\frac{1}{2}\bm{\beta}^{\mathsf T}\bm{Q}_\beta\bm{\beta}
+\bm{\beta}^{\mathsf T}\bm{h}_\beta+\mathrm{const}\\
&= -\frac{1}{2}\left(\bm{\beta}-\bm{\mu}_\beta\right)^{\mathsf T}
\bm{Q}_\beta\left(\bm{\beta}-\bm{\mu}_\beta\right)+\mathrm{const},
\end{aligned}
\label{eq:complete-square}
\end{equation}
where $\bm{\mu}_\beta = \bm{Q}_\beta^{-1}\bm{h}_\beta$, $\bm{Q}_\beta$ and $\bm{h}_\beta$ are given in \eqref{eq:beta-gibbs}, and $\mathrm{const}$ denotes terms independent of $\bm{\beta}$ that may differ between lines. For any nonzero $\bm{v}\in\R^p$,
\begin{equation}
\bm{v}^{\mathsf T}\bm{Q}_\beta\bm{v}
=\bm{v}^{\mathsf T}\bm{V}_\beta^{-1}\bm{v}
+\sum_{i=1}^{n}\left(\bm{H}_i\bm{v}\right)^{\mathsf T}
\bm{\Sigma}^{-1}\left(\bm{H}_i\bm{v}\right)>0.
\end{equation}
This establishes both the Gaussian conditional and its nonsingular covariance. It does not require each $\bm{H}_i$ to have full column rank.

For computation, let $\bm{Q}_\beta=\bm{T}_\beta\bm{T}_\beta^{\mathsf T}$ be its lower triangular Cholesky factorization. Solve $\bm{Q}_\beta\bm{\mu}_\beta=\bm{h}_\beta$ by two triangular solves and generate
\begin{equation}
\bm{\beta}^{\mathrm{new}}
=\bm{\mu}_\beta+\bm{T}_\beta^{-\mathsf T}\bm{e},
\qquad\bm{e}\sim\Normal_p\left(\bm{0},\bm{I}_p\right).
\label{eq:precision-sampling}
\end{equation}
The covariance of the added term in \eqref{eq:precision-sampling} is $\bm{T}_\beta^{-\mathsf T}\bm{T}_\beta^{-1}=\bm{Q}_\beta^{-1}$.

For the application design in \eqref{eq:bvrt-design}, the required matrix is
\begin{equation}
\bm{H}_i=
\begin{pmatrix}
1&0&E_i&A_i&M_i\\
0&1&0&0&0
\end{pmatrix}.
\label{eq:application-h}
\end{equation}
Thus $\bm{c}_i$ contains the participant's total intercept and time slope. Equation \eqref{eq:application-h} verifies the compatibility condition directly, including the fixed effects of education, entry age and sex.

\section{Covariance densities and Jacobians}
\label{app:covariance}
\setcounter{equation}{0}

\subsection{Separation prior in unconstrained coordinates}

Continue to suppress the fixed quantile index $\tau$. For $q=2$, independent half-normal priors and the $\LKJ_2\left(\nu\right)$ prior give the density
\begin{equation}
\pi\left(\sigma_1,\sigma_2,\rho\right)
=\frac{2}{\pi s_1s_2\,\mathrm{B}\left(1/2,\nu\right)}
\exp\left\{-\frac{1}{2}\left(\frac{\sigma_1^2}{s_1^2}+\frac{\sigma_2^2}{s_2^2}\right)\right\}
\left(1-\rho^2\right)^{\nu-1},
\label{eq:normalized-separation-prior}
\end{equation}
for $\sigma_1,\sigma_2>0$ and $-1<\rho<1$, where $\mathrm{B}$ denotes the beta function. This density is with respect to $\mathrm{d}\sigma_1\,\mathrm{d}\sigma_2\,\mathrm{d}\rho$. The inverse of \eqref{eq:covariance-coordinates} has derivative matrix
\begin{equation}
\frac{\partial\left(\sigma_1,\sigma_2,\rho\right)}
{\partial\left(\lambda_1,\lambda_2,\zeta\right)}
=\diag\left(\sigma_1,\sigma_2,1-\rho^2\right).
\end{equation}
Multiplying \eqref{eq:normalized-separation-prior} by its determinant and taking logarithms gives \eqref{eq:transformed-covariance-prior}. No additional transformation to matrix-entry coordinates is involved, because the prior was specified directly on SDs and correlation.

The covariance and its lower triangular Cholesky factor are
\begin{equation}
\bm{\Sigma}=
\begin{pmatrix}
\sigma_1^2&\rho\sigma_1\sigma_2\\
\rho\sigma_1\sigma_2&\sigma_2^2
\end{pmatrix},
\qquad
\bm{C}=
\begin{pmatrix}
\sigma_1&0\\
\rho\sigma_2&\sigma_2\sqrt{1-\rho^2}
\end{pmatrix}.
\label{eq:explicit-cholesky}
\end{equation}
In particular, $\det\bm{\Sigma}=\sigma_1^2\sigma_2^2\left(1-\rho^2\right)$ and $\det\bm{C}>0$. Writing $S_{k\ell}$ for the entries of $\bm{S}_b$, the centered log target in \eqref{eq:centered-covariance-target} can equivalently be evaluated as
\begin{equation}
t_{\mathrm{C}}\left(\bm{\psi}\right)
=g\left(\bm{\psi}\right)-n\left(\lambda_1+\lambda_2\right)
-\frac{n}{2}\log\left(1-\rho^2\right)-\frac{1}{2\left(1-\rho^2\right)}
\left(\frac{S_{11}}{\sigma_1^2}-\frac{2\rho S_{12}}{\sigma_1\sigma_2}
+\frac{S_{22}}{\sigma_2^2}\right).
\label{eq:centered-expanded}
\end{equation}
Here, the data likelihood contributes no term in $\bm{\psi}$ because both $\bm{\beta}$ and $\bm{b}_{1:n}$ are fixed.

\subsection{Whitened coordinates}

For fixed $\bm{\beta}$, consider the joint transformation
$\left(\bm{u}_{1:n},\bm{\psi}\right)\mapsto
\left(\bm{b}_{1:n},\bm{\psi}\right)$ with $\bm{b}_i=\bm{C}\left(\bm{\psi}\right)\bm{u}_i$.
Although $\bm{b}_i$ depends on $\bm{\psi}$, the full derivative matrix is block triangular. Its determinant is $\det\left\{\bm{C}\left(\bm{\psi}\right)\right\}^{n}$. For each participant,
\begin{equation}
\begin{aligned}
&\varphi_q\left(\bm{C}\bm{u}_i;\bm{0},\bm{\Sigma}\right)\det\bm{C}\\
&\qquad=\left(2\pi\right)^{-q/2}
\det\left(\bm{\Sigma}\right)^{-1/2}\det\bm{C}
\exp\left(-\frac{1}{2}\bm{u}_i^{\mathsf T}\bm{u}_i\right)
=\varphi_q\left(\bm{u}_i;\bm{0},\bm{I}_q\right).
\end{aligned}
\label{eq:whitening-cancellation}
\end{equation}
Thus the posterior density in these coordinates is proportional to
\begin{equation}
\begin{split}
&\pi\left(\bm{\beta}\right)\pi\left(\bm{\gamma}\right)
\exp\left\{g\left(\bm{\psi}\right)\right\}
\prod_{i=1}^{n}\varphi_q\left(\bm{u}_i;\bm{0},\bm{I}_q\right)\\
&\qquad\times
\exp\left\{\sum_{i=1}^{n}\ell_i\left(
\bm{H}_i\bm{\beta}+\bm{C}\left(\bm{\psi}\right)\bm{u}_i,
\bm{\gamma}\right)\right\}.
\end{split}
\label{eq:whitened-posterior}
\end{equation}
At fixed $\bm{u}_{1:n}$, $\bm{\beta}$ and $\bm{\gamma}$, taking logarithms of \eqref{eq:whitened-posterior} and retaining only the terms that depend on $\bm{\psi}$ gives the log target in \eqref{eq:noncentered-covariance-target}. The Jacobian for the covariance parameter transformation is already included in $g$ and remains necessary; it is distinct from the Jacobian for the participant effects canceled in \eqref{eq:whitening-cancellation}.

\section{Invariance of sampling sweep}
\label{app:invariance}
\setcounter{equation}{0}

Let $\Pi$ denote the proper posterior probability measure in total-coefficient coordinates. The Gaussian draw in \eqref{eq:beta-gibbs} is a full conditional update. The participant and shape steps use elliptical slice kernels for the conditional densities in \eqref{eq:effect-conditional} and \eqref{eq:shape-conditional}; these kernels preserve a Gaussian prior multiplied by a likelihood \citep{murray2010elliptical}. The scalar stepping-out and shrinkage updates preserve the respective one-coordinate conditionals with log targets $t_{\mathrm{C}}$ or $t_{\mathrm{N}}$ \citep{neal2003slice}. The implemented widths are fixed; a numerical safeguard terminates with an error rather than silently truncating a transition.

To justify the coordinate changes, let $T$ be an invertible map and let $\Pi_T$ be the distribution of the transformed state $\bm{v}=T\left(\bm{s}\right)$ when the original sampler state $\bm{s}\sim\Pi$. If a kernel $K_T$ preserves $\Pi_T$, transforming to the new coordinates, applying that kernel, and transforming back gives
\begin{equation}
P_T\left(\bm{s},A\right)
=K_T\left(T\left(\bm{s}\right),T\left(A\right)\right)
\end{equation}
for a measurable set $A$ of original states. Its invariance follows from
\begin{equation}
\begin{aligned}
\int P_T\left(\bm{s},A\right)\,\Pi\left(\mathrm{d}\bm{s}\right)
&=\int K_T\left(\bm{v},T\left(A\right)\right)\,\Pi_T\left(\mathrm{d}\bm{v}\right)\\
&=\Pi_T\left(T\left(A\right)\right)=\Pi\left(A\right).
\end{aligned}
\label{eq:transformed-kernel-invariance}
\end{equation}
The total-coefficient and whitening transformations satisfy these conditions, with their densities derived in \appref{app:gaussian} and \appref{app:covariance}. Each full conditional or slice update, expressed back in the common coordinates, therefore preserves $\Pi$. If $P_1,\ldots,P_J$ denote the $J$ constituent kernels of a full sweep in the order of \mainalgref{alg:sampler}, then
\begin{equation}
\Pi P_1P_2\cdots P_J=\Pi.
\label{eq:sweep-invariance}
\end{equation}
This argument justifies the posterior target of the interwoven sweep. It is not a rate-of-convergence result and does not imply that a finite chain has reached equilibrium. In particular, no claim about improved mixing follows from invariance alone.

\StartSupplement

\begin{landscape}

\section{Posterior parameter summaries}
\label{supp:parameter-summaries}

\begingroup
\small
\setlength{\tabcolsep}{4pt}
\renewcommand{\arraystretch}{1.0}
\setlength{\LTleft}{\fill}
\setlength{\LTright}{\fill}
\begin{longtable}{@{}
    >{\arraybackslash}m{1.8cm}
    >{\arraybackslash}m{4.0cm}
    *{5}{>{\centering\arraybackslash}m{\dimexpr(\linewidth-5.8cm-12\tabcolsep)/5\relax}}
    @{}}
\caption{Posterior summaries for the BVRT quantile mixed models.}\label{tab:bvrt-parameters}\\
\toprule
\textbf{Parameter} & \textbf{Description} & $\bm{\tau = 0.10}$ & $\bm{\tau = 0.25}$ & $\bm{\tau = 0.50}$ & $\bm{\tau = 0.75}$ & $\bm{\tau = 0.90}$ \\
\midrule
\endfirsthead
\multicolumn{7}{c}{\tablename\ \thetable{} (continued)} \\
\toprule
\textbf{Parameter} & \textbf{Description} & $\bm{\tau = 0.10}$ & $\bm{\tau = 0.25}$ & $\bm{\tau = 0.50}$ & $\bm{\tau = 0.75}$ & $\bm{\tau = 0.90}$ \\
\midrule
\endhead
\midrule
\multicolumn{7}{r}{\textit{Continued on next page}} \\
\endfoot
\bottomrule
\multicolumn{7}{@{}p{0.96\linewidth}@{}}{\footnotesize \textit{Notes:} Entries are posterior medians, with 95\% equal-tailed credible intervals in brackets, using all 40,000 post-warmup draws per fit. Each column represents a separately fitted model. The $\beta$ coefficients act on the logit of the scaled quantile anchor; the $\gamma$ coefficients act on $\log\phi$. Reference categories are no primary school graduation and female sex. Entry age is centered at the participant-level sample mean. Participant-effect SDs are on the logit predictor scale, with the slope expressed per five years.} \\
\endlastfoot
\multicolumn{7}{@{}l}{\textit{Quantile regression}} \\*
\addlinespace[0.25em]
$\beta_{0,\tau}$ & Intercept & \begin{tabular}[c]{@{}c@{}}$-0.157$\\$\left[-0.303;\,-0.015\right]$\end{tabular} & \begin{tabular}[c]{@{}c@{}}$0.215$\\$\left[0.088;\,0.341\right]$\end{tabular} & \begin{tabular}[c]{@{}c@{}}$0.598$\\$\left[0.484;\,0.714\right]$\end{tabular} & \begin{tabular}[c]{@{}c@{}}$0.966$\\$\left[0.853;\,1.079\right]$\end{tabular} & \begin{tabular}[c]{@{}c@{}}$1.290$\\$\left[1.166;\,1.415\right]$\end{tabular} \\[0.3em]
$\beta_{t,\tau}$ & Time (per 5 years) & \begin{tabular}[c]{@{}c@{}}$-0.189$\\$\left[-0.229;\,-0.151\right]$\end{tabular} & \begin{tabular}[c]{@{}c@{}}$-0.178$\\$\left[-0.215;\,-0.143\right]$\end{tabular} & \begin{tabular}[c]{@{}c@{}}$-0.168$\\$\left[-0.203;\,-0.134\right]$\end{tabular} & \begin{tabular}[c]{@{}c@{}}$-0.161$\\$\left[-0.194;\,-0.129\right]$\end{tabular} & \begin{tabular}[c]{@{}c@{}}$-0.156$\\$\left[-0.188;\,-0.126\right]$\end{tabular} \\[0.3em]
$\beta_{E,\tau}$ & Primary school graduation & \begin{tabular}[c]{@{}c@{}}$0.596$\\$\left[0.446;\,0.750\right]$\end{tabular} & \begin{tabular}[c]{@{}c@{}}$0.578$\\$\left[0.445;\,0.711\right]$\end{tabular} & \begin{tabular}[c]{@{}c@{}}$0.581$\\$\left[0.457;\,0.703\right]$\end{tabular} & \begin{tabular}[c]{@{}c@{}}$0.600$\\$\left[0.478;\,0.723\right]$\end{tabular} & \begin{tabular}[c]{@{}c@{}}$0.632$\\$\left[0.496;\,0.767\right]$\end{tabular} \\[0.3em]
$\beta_{A,\tau}$ & Entry age (per 10 years) & \begin{tabular}[c]{@{}c@{}}$-0.521$\\$\left[-0.624;\,-0.417\right]$\end{tabular} & \begin{tabular}[c]{@{}c@{}}$-0.486$\\$\left[-0.584;\,-0.390\right]$\end{tabular} & \begin{tabular}[c]{@{}c@{}}$-0.457$\\$\left[-0.550;\,-0.365\right]$\end{tabular} & \begin{tabular}[c]{@{}c@{}}$-0.437$\\$\left[-0.525;\,-0.349\right]$\end{tabular} & \begin{tabular}[c]{@{}c@{}}$-0.421$\\$\left[-0.507;\,-0.337\right]$\end{tabular} \\[0.3em]
$\beta_{M,\tau}$ & Male sex & \begin{tabular}[c]{@{}c@{}}$0.156$\\$\left[0.030;\,0.282\right]$\end{tabular} & \begin{tabular}[c]{@{}c@{}}$0.146$\\$\left[0.031;\,0.262\right]$\end{tabular} & \begin{tabular}[c]{@{}c@{}}$0.140$\\$\left[0.029;\,0.248\right]$\end{tabular} & \begin{tabular}[c]{@{}c@{}}$0.133$\\$\left[0.029;\,0.239\right]$\end{tabular} & \begin{tabular}[c]{@{}c@{}}$0.131$\\$\left[0.029;\,0.232\right]$\end{tabular} \\[0.3em]
\addlinespace[0.4em]
\multicolumn{7}{@{}l}{\textit{Shape regression}} \\*
\addlinespace[0.25em]
$\gamma_{0,\tau}$ & Intercept & \begin{tabular}[c]{@{}c@{}}$1.469$\\$\left[1.413;\,1.524\right]$\end{tabular} & \begin{tabular}[c]{@{}c@{}}$1.214$\\$\left[1.143;\,1.281\right]$\end{tabular} & \begin{tabular}[c]{@{}c@{}}$0.923$\\$\left[0.827;\,1.010\right]$\end{tabular} & \begin{tabular}[c]{@{}c@{}}$0.626$\\$\left[0.509;\,0.738\right]$\end{tabular} & \begin{tabular}[c]{@{}c@{}}$0.350$\\$\left[0.203;\,0.490\right]$\end{tabular} \\[0.3em]
$\gamma_{E,\tau}$ & Primary school graduation & \begin{tabular}[c]{@{}c@{}}$-0.072$\\$\left[-0.134;\,-0.009\right]$\end{tabular} & \begin{tabular}[c]{@{}c@{}}$-0.098$\\$\left[-0.176;\,-0.020\right]$\end{tabular} & \begin{tabular}[c]{@{}c@{}}$-0.133$\\$\left[-0.234;\,-0.027\right]$\end{tabular} & \begin{tabular}[c]{@{}c@{}}$-0.178$\\$\left[-0.311;\,-0.043\right]$\end{tabular} & \begin{tabular}[c]{@{}c@{}}$-0.228$\\$\left[-0.392;\,-0.061\right]$\end{tabular} \\[0.3em]
\addlinespace[0.4em]
\multicolumn{7}{@{}l}{\textit{Participant effects}} \\*
\addlinespace[0.25em]
$\sigma_{1,\tau}$ & Random intercept SD & \begin{tabular}[c]{@{}c@{}}$0.630$\\$\left[0.558;\,0.706\right]$\end{tabular} & \begin{tabular}[c]{@{}c@{}}$0.592$\\$\left[0.524;\,0.663\right]$\end{tabular} & \begin{tabular}[c]{@{}c@{}}$0.559$\\$\left[0.495;\,0.628\right]$\end{tabular} & \begin{tabular}[c]{@{}c@{}}$0.536$\\$\left[0.473;\,0.600\right]$\end{tabular} & \begin{tabular}[c]{@{}c@{}}$0.520$\\$\left[0.459;\,0.584\right]$\end{tabular} \\[0.3em]
$\sigma_{2,\tau}$ & Random slope SD & \begin{tabular}[c]{@{}c@{}}$0.162$\\$\left[0.108;\,0.214\right]$\end{tabular} & \begin{tabular}[c]{@{}c@{}}$0.151$\\$\left[0.096;\,0.199\right]$\end{tabular} & \begin{tabular}[c]{@{}c@{}}$0.141$\\$\left[0.089;\,0.188\right]$\end{tabular} & \begin{tabular}[c]{@{}c@{}}$0.135$\\$\left[0.084;\,0.178\right]$\end{tabular} & \begin{tabular}[c]{@{}c@{}}$0.130$\\$\left[0.079;\,0.173\right]$\end{tabular} \\[0.3em]
$\rho_\tau$ & Intercept-slope correlation & \begin{tabular}[c]{@{}c@{}}$-0.531$\\$\left[-0.695;\,-0.287\right]$\end{tabular} & \begin{tabular}[c]{@{}c@{}}$-0.536$\\$\left[-0.695;\,-0.284\right]$\end{tabular} & \begin{tabular}[c]{@{}c@{}}$-0.536$\\$\left[-0.701;\,-0.290\right]$\end{tabular} & \begin{tabular}[c]{@{}c@{}}$-0.543$\\$\left[-0.704;\,-0.287\right]$\end{tabular} & \begin{tabular}[c]{@{}c@{}}$-0.548$\\$\left[-0.704;\,-0.295\right]$\end{tabular} \\[0.3em]
\end{longtable}
\endgroup
\end{landscape}

\section{Posterior predictive score frequencies}
\label{supp:score-frequencies}

\begin{figure}[!htbp]
\centering
\includegraphics[width=\textwidth]{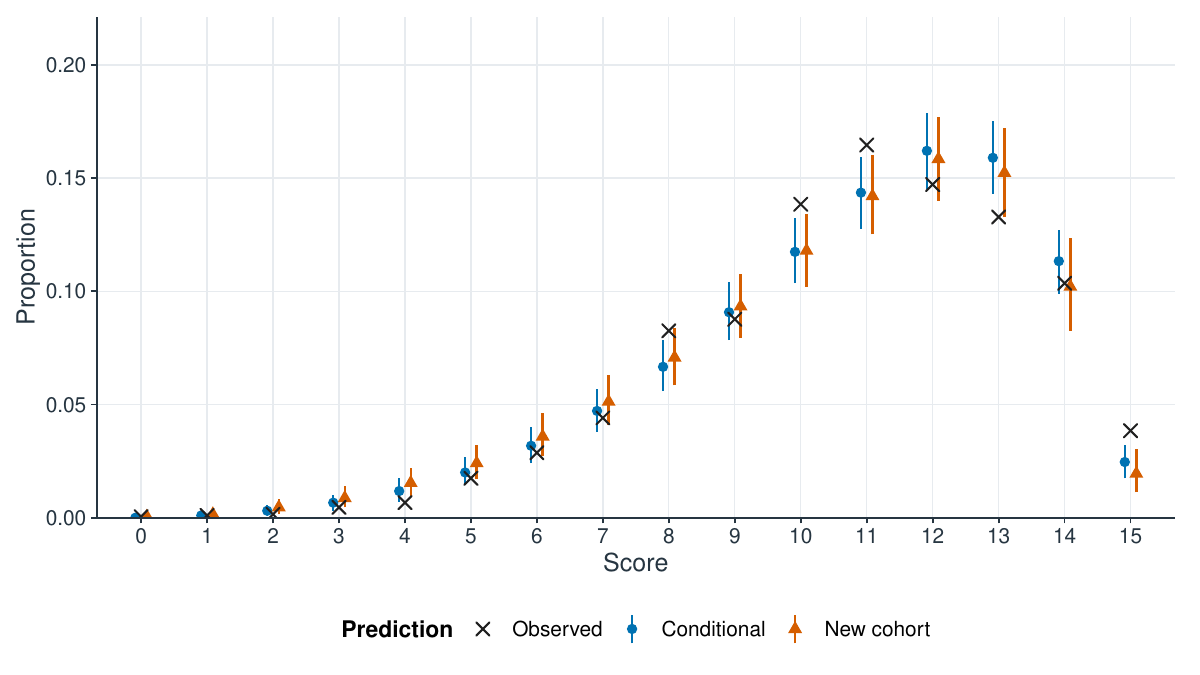}
\caption{Posterior predictive BVRT score frequencies overall for the model fitted at $\tau = 0.10$. Black crosses show observed proportions, blue circles show conditional predictive medians using the sampled effects of existing participants, and orange triangles show predictive medians using newly generated participant effects. Vertical lines indicate pointwise 95\% posterior predictive intervals. Proportions pool all recorded assessments. Both prediction schemes retain the observed covariates, assessment times, and pattern of recorded outcomes, with one shared effect vector per participant. All post-warmup draws are used.}
\label{fig:bvrt-frequencies-q10}
\end{figure}
\clearpage

\begin{figure}[p]
\centering
\captionsetup{font=small}
\begin{subfigure}{\textwidth}
\centering
\includegraphics[width=\linewidth]{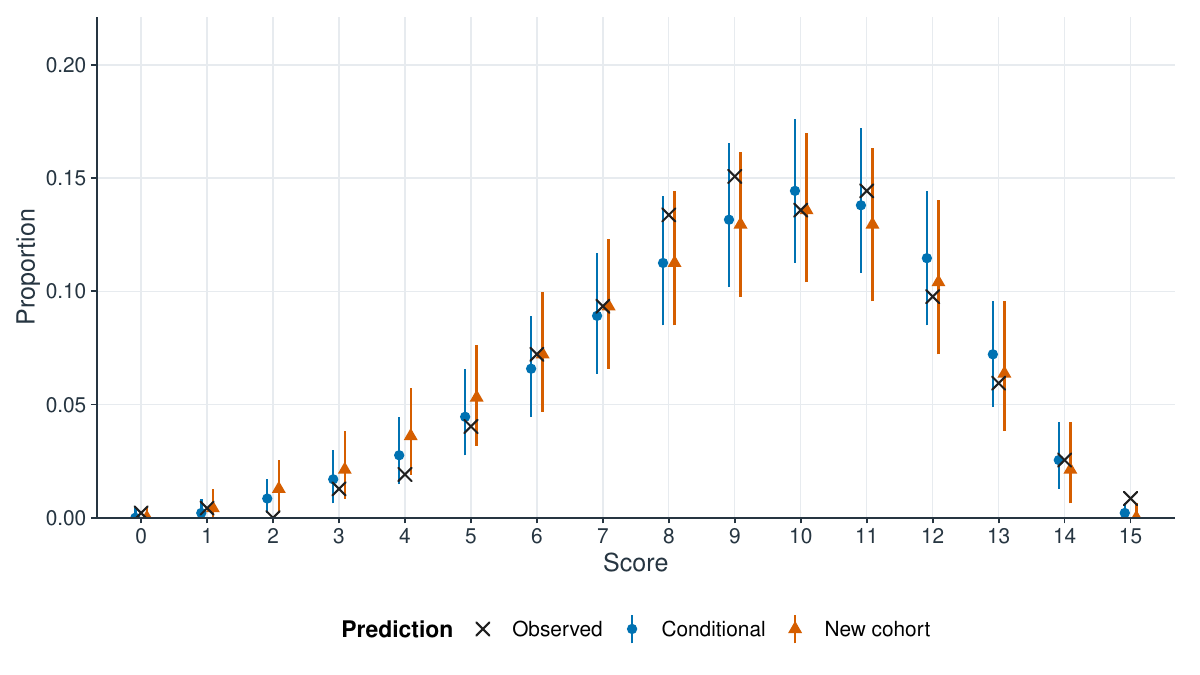}
\caption{No primary school graduation}
\end{subfigure}

\smallskip

\begin{subfigure}{\textwidth}
\centering
\includegraphics[width=\linewidth]{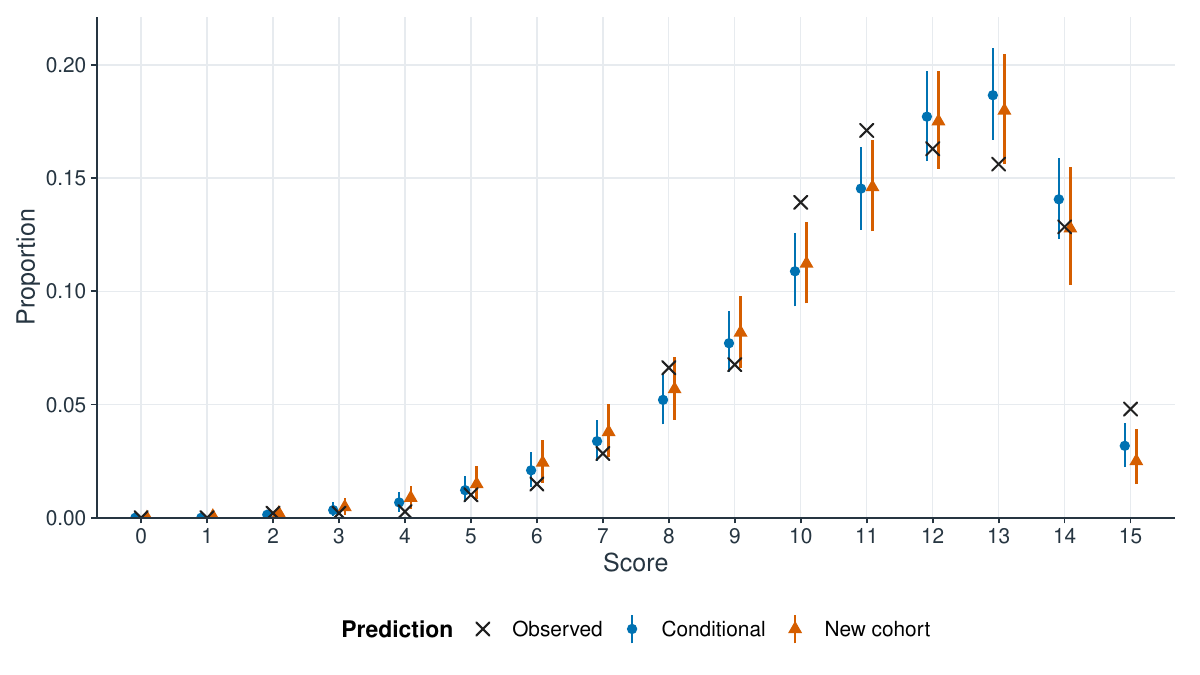}
\caption{Primary school graduates}
\end{subfigure}
\caption{Posterior predictive BVRT score frequencies by education for the model fitted at $\tau = 0.10$. Black crosses show observed proportions; blue circles and orange triangles show conditional and new-cohort predictive medians, respectively. Vertical lines indicate pointwise 95\% posterior predictive intervals. Proportions pool recorded assessments within each group. Prediction schemes and use of all post-warmup draws follow \suppfigref{fig:bvrt-frequencies-q10}.}
\label{fig:bvrt-frequencies-education-q10}
\end{figure}
\clearpage

\begin{figure}[p]
\centering
\captionsetup{font=small}
\begin{subfigure}{\textwidth}
\centering
\includegraphics[width=\linewidth]{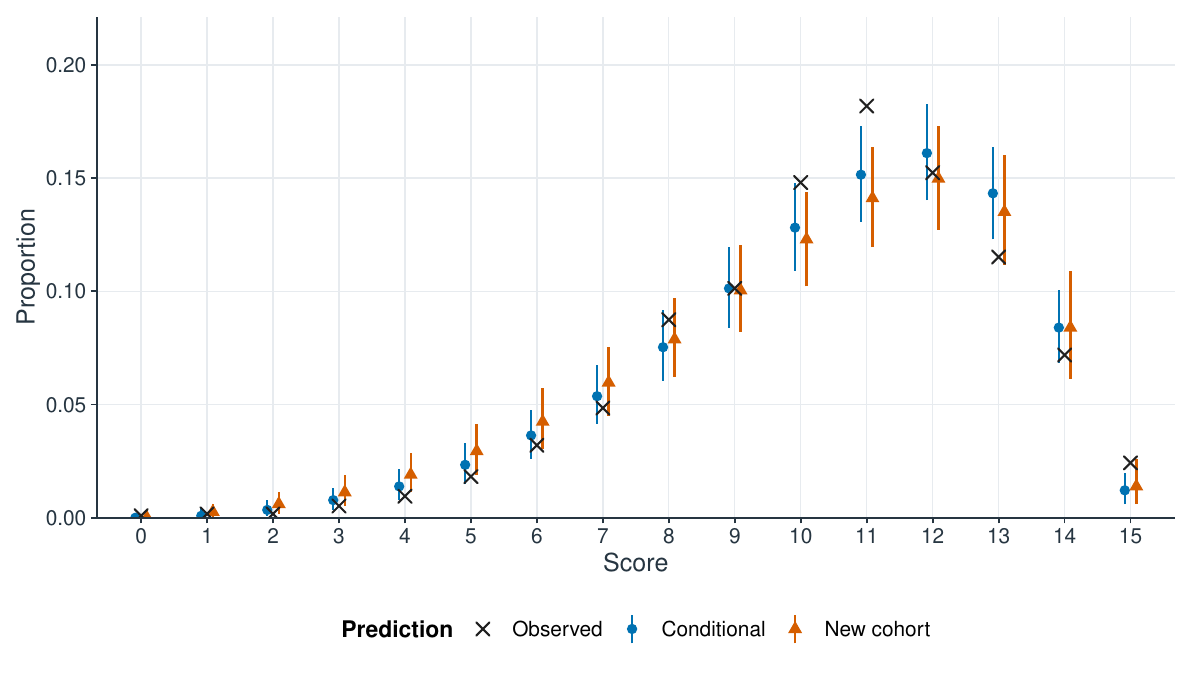}
\caption{Female}
\end{subfigure}

\smallskip

\begin{subfigure}{\textwidth}
\centering
\includegraphics[width=\linewidth]{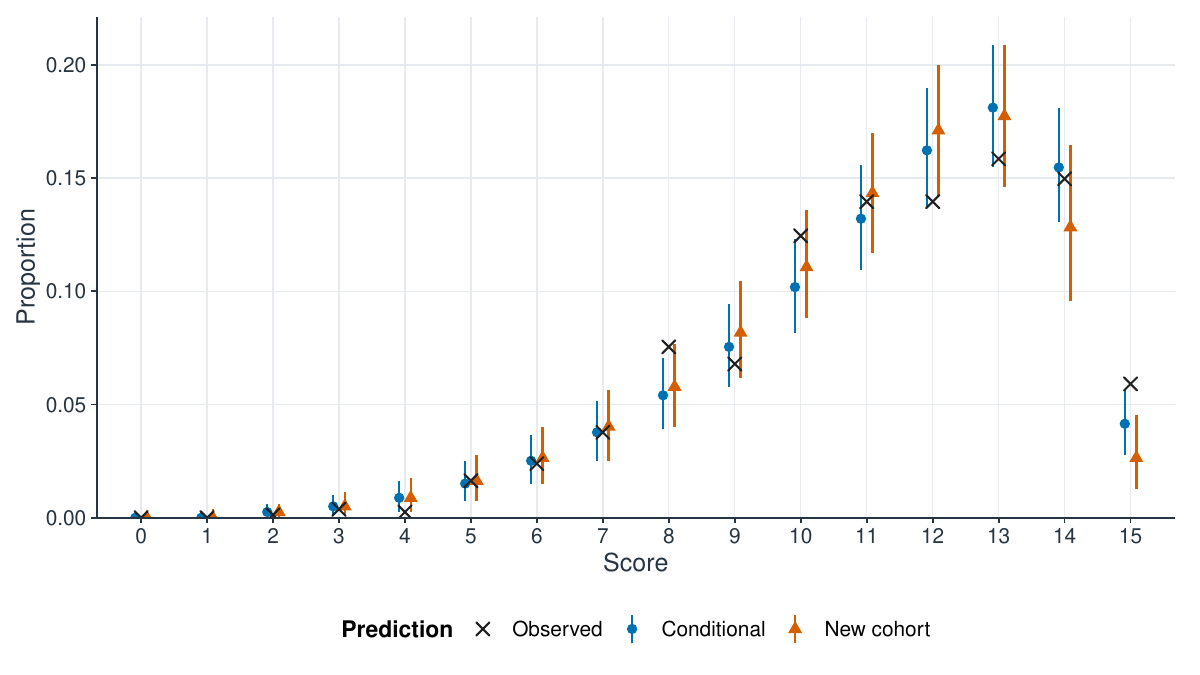}
\caption{Male}
\end{subfigure}
\caption{Posterior predictive BVRT score frequencies by sex for the model fitted at $\tau = 0.10$. Black crosses show observed proportions; blue circles and orange triangles show conditional and new-cohort predictive medians, respectively. Vertical lines indicate pointwise 95\% posterior predictive intervals. Proportions pool recorded assessments within each group. Prediction schemes and use of all post-warmup draws follow \suppfigref{fig:bvrt-frequencies-q10}.}
\label{fig:bvrt-frequencies-sex-q10}
\end{figure}
\clearpage

\begin{figure}[!htbp]
\centering
\includegraphics[width=\textwidth]{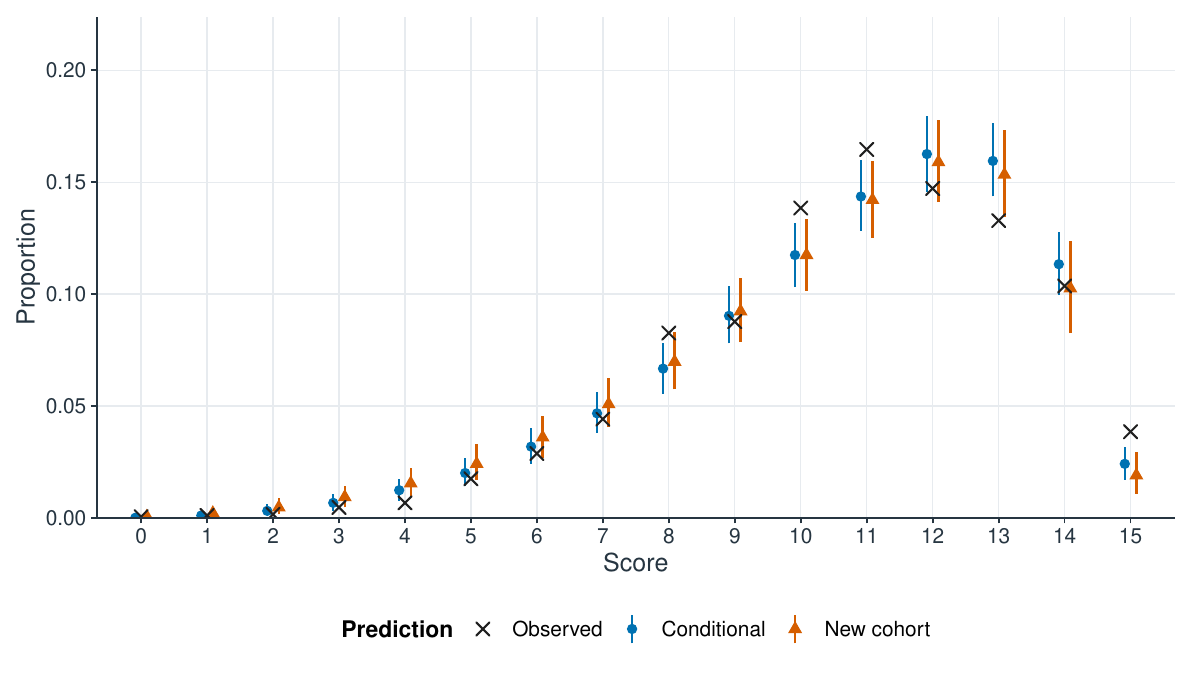}
\caption{Posterior predictive BVRT score frequencies overall for the model fitted at $\tau = 0.25$. Black crosses show observed proportions, blue circles show conditional predictive medians using the sampled effects of existing participants, and orange triangles show predictive medians using newly generated participant effects. Vertical lines indicate pointwise 95\% posterior predictive intervals. Proportions pool all recorded assessments. Both prediction schemes retain the observed covariates, assessment times, and pattern of recorded outcomes, with one shared effect vector per participant. All post-warmup draws are used.}
\label{fig:bvrt-frequencies-q25}
\end{figure}
\clearpage

\begin{figure}[p]
\centering
\captionsetup{font=small}
\begin{subfigure}{\textwidth}
\centering
\includegraphics[width=\linewidth]{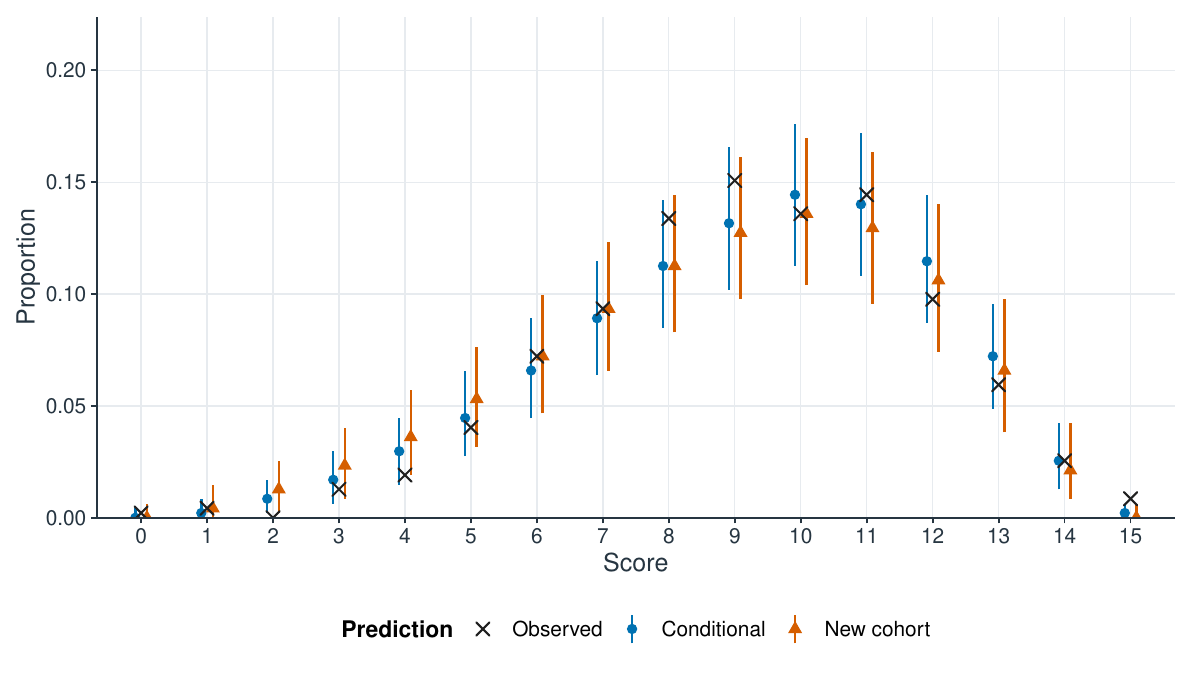}
\caption{No primary school graduation}
\end{subfigure}

\smallskip

\begin{subfigure}{\textwidth}
\centering
\includegraphics[width=\linewidth]{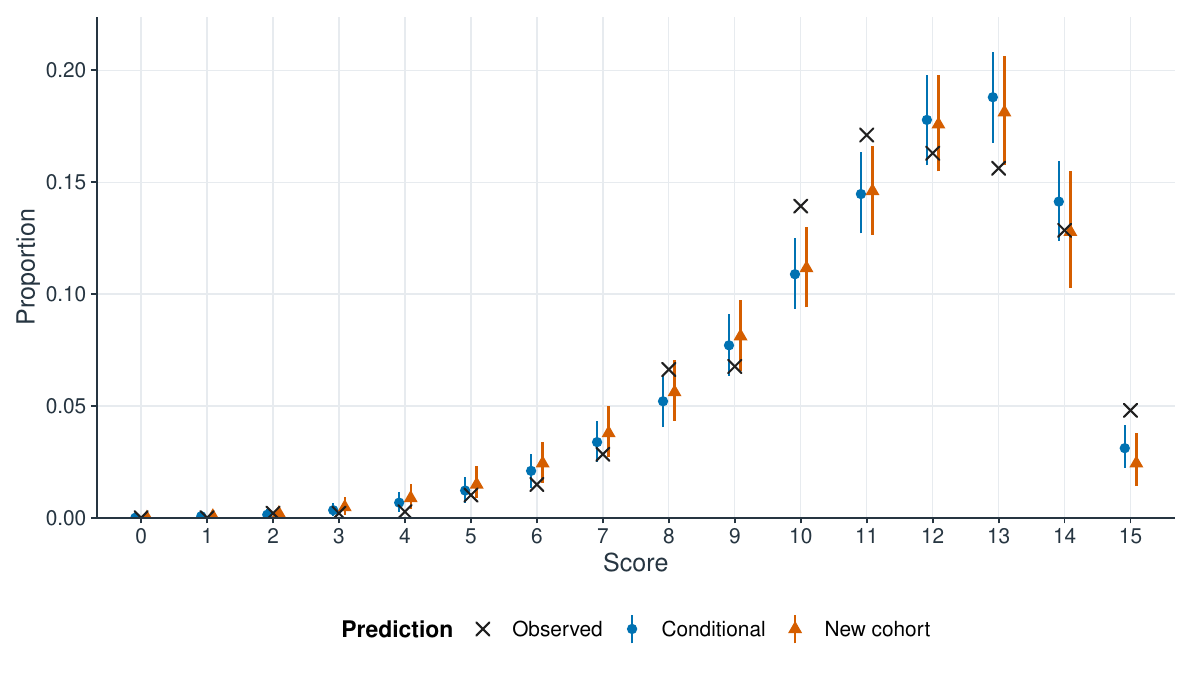}
\caption{Primary school graduates}
\end{subfigure}
\caption{Posterior predictive BVRT score frequencies by education for the model fitted at $\tau = 0.25$. Black crosses show observed proportions; blue circles and orange triangles show conditional and new-cohort predictive medians, respectively. Vertical lines indicate pointwise 95\% posterior predictive intervals. Proportions pool recorded assessments within each group. Prediction schemes and use of all post-warmup draws follow \suppfigref{fig:bvrt-frequencies-q25}.}
\label{fig:bvrt-frequencies-education-q25}
\end{figure}
\clearpage

\begin{figure}[p]
\centering
\captionsetup{font=small}
\begin{subfigure}{\textwidth}
\centering
\includegraphics[width=\linewidth]{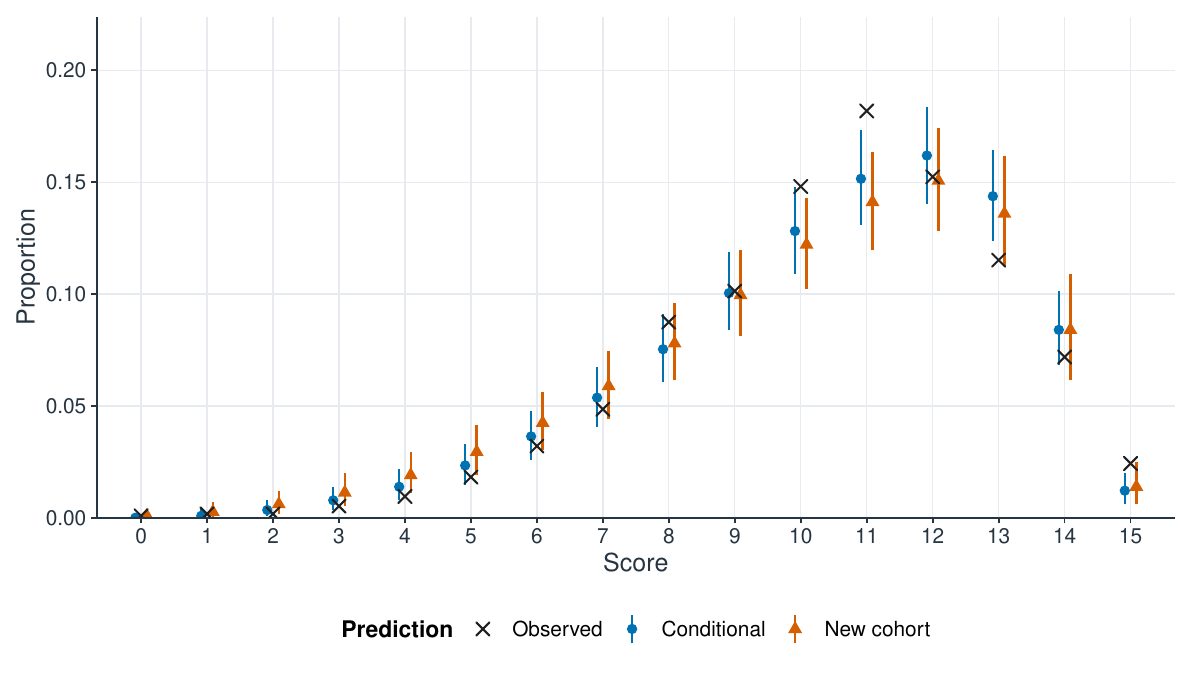}
\caption{Female}
\end{subfigure}

\smallskip

\begin{subfigure}{\textwidth}
\centering
\includegraphics[width=\linewidth]{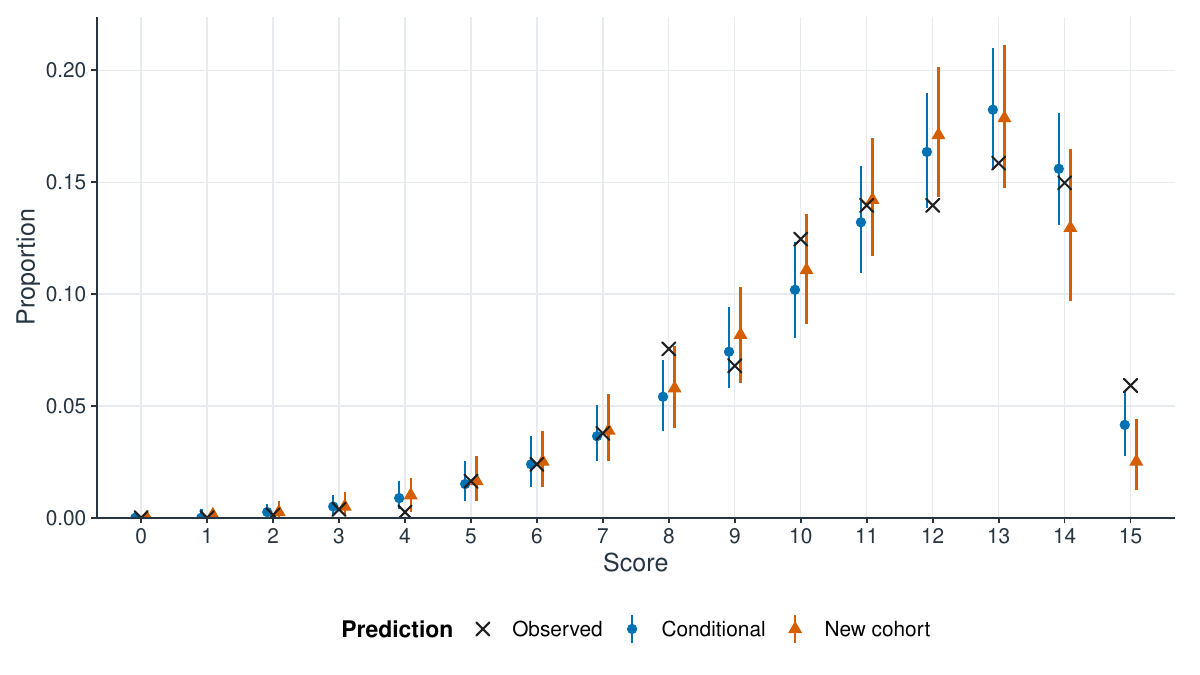}
\caption{Male}
\end{subfigure}
\caption{Posterior predictive BVRT score frequencies by sex for the model fitted at $\tau = 0.25$. Black crosses show observed proportions; blue circles and orange triangles show conditional and new-cohort predictive medians, respectively. Vertical lines indicate pointwise 95\% posterior predictive intervals. Proportions pool recorded assessments within each group. Prediction schemes and use of all post-warmup draws follow \suppfigref{fig:bvrt-frequencies-q25}.}
\label{fig:bvrt-frequencies-sex-q25}
\end{figure}
\clearpage

\begin{figure}[!htbp]
\centering
\includegraphics[width=\textwidth]{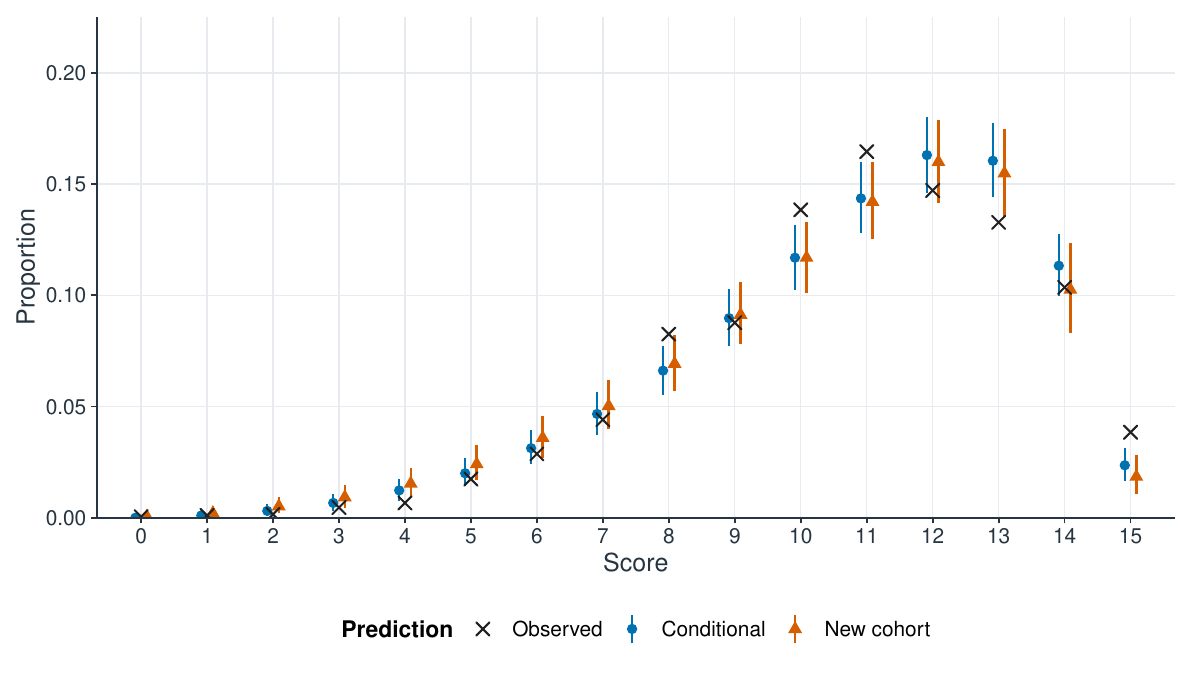}
\caption{Posterior predictive BVRT score frequencies overall for the model fitted at $\tau = 0.50$. Black crosses show observed proportions, blue circles show conditional predictive medians using the sampled effects of existing participants, and orange triangles show predictive medians using newly generated participant effects. Vertical lines indicate pointwise 95\% posterior predictive intervals. Proportions pool all recorded assessments. Both prediction schemes retain the observed covariates, assessment times, and pattern of recorded outcomes, with one shared effect vector per participant. All post-warmup draws are used.}
\label{fig:bvrt-frequencies-q50}
\end{figure}
\clearpage

\begin{figure}[p]
\centering
\captionsetup{font=small}
\begin{subfigure}{\textwidth}
\centering
\includegraphics[width=\linewidth]{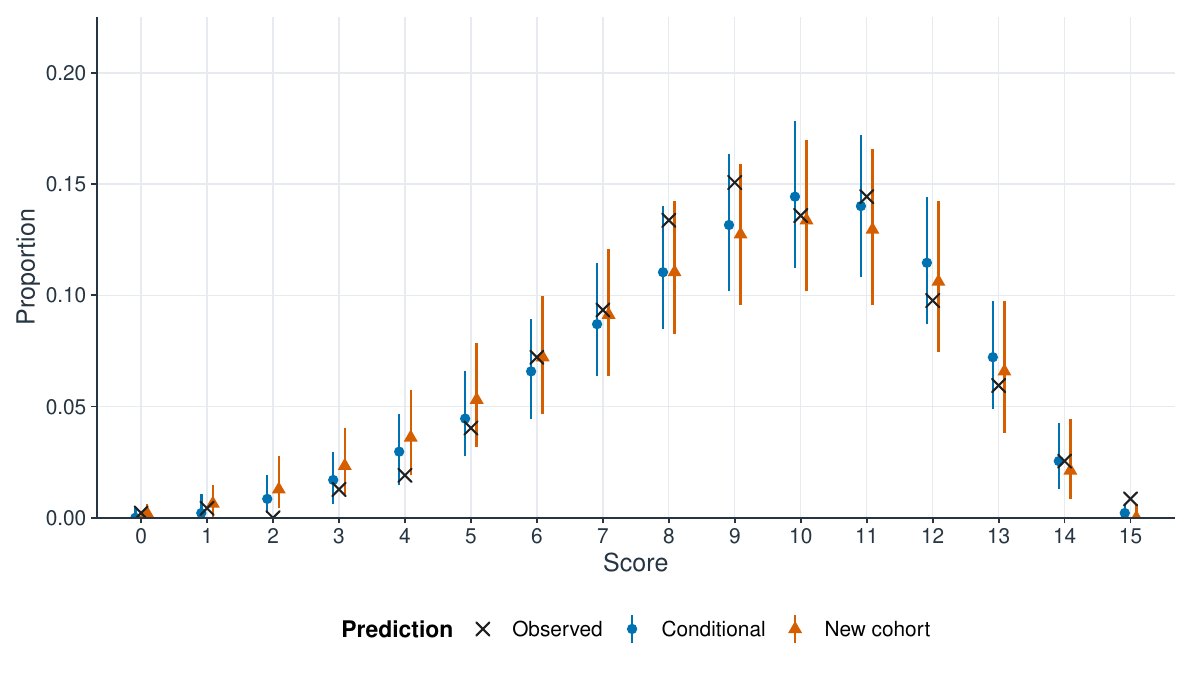}
\caption{No primary school graduation}
\end{subfigure}

\smallskip

\begin{subfigure}{\textwidth}
\centering
\includegraphics[width=\linewidth]{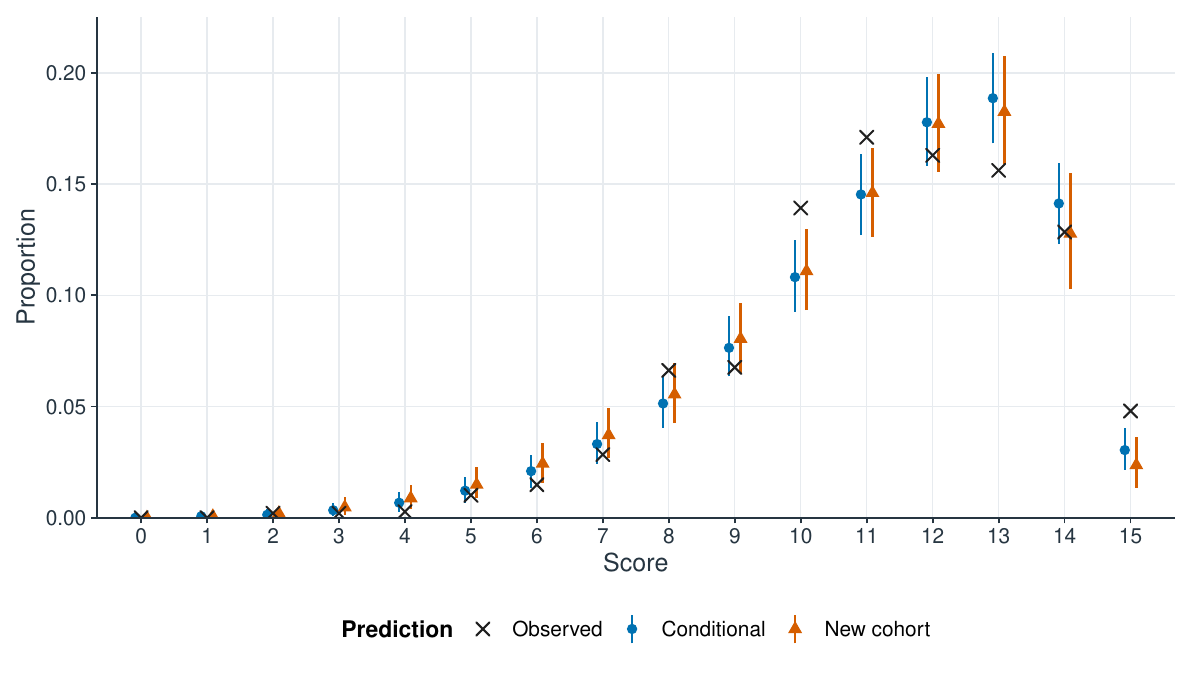}
\caption{Primary school graduates}
\end{subfigure}
\caption{Posterior predictive BVRT score frequencies by education for the model fitted at $\tau = 0.50$. Black crosses show observed proportions; blue circles and orange triangles show conditional and new-cohort predictive medians, respectively. Vertical lines indicate pointwise 95\% posterior predictive intervals. Proportions pool recorded assessments within each group. Prediction schemes and use of all post-warmup draws follow \suppfigref{fig:bvrt-frequencies-q50}.}
\label{fig:bvrt-frequencies-education-q50}
\end{figure}
\clearpage

\begin{figure}[p]
\centering
\captionsetup{font=small}
\begin{subfigure}{\textwidth}
\centering
\includegraphics[width=\linewidth]{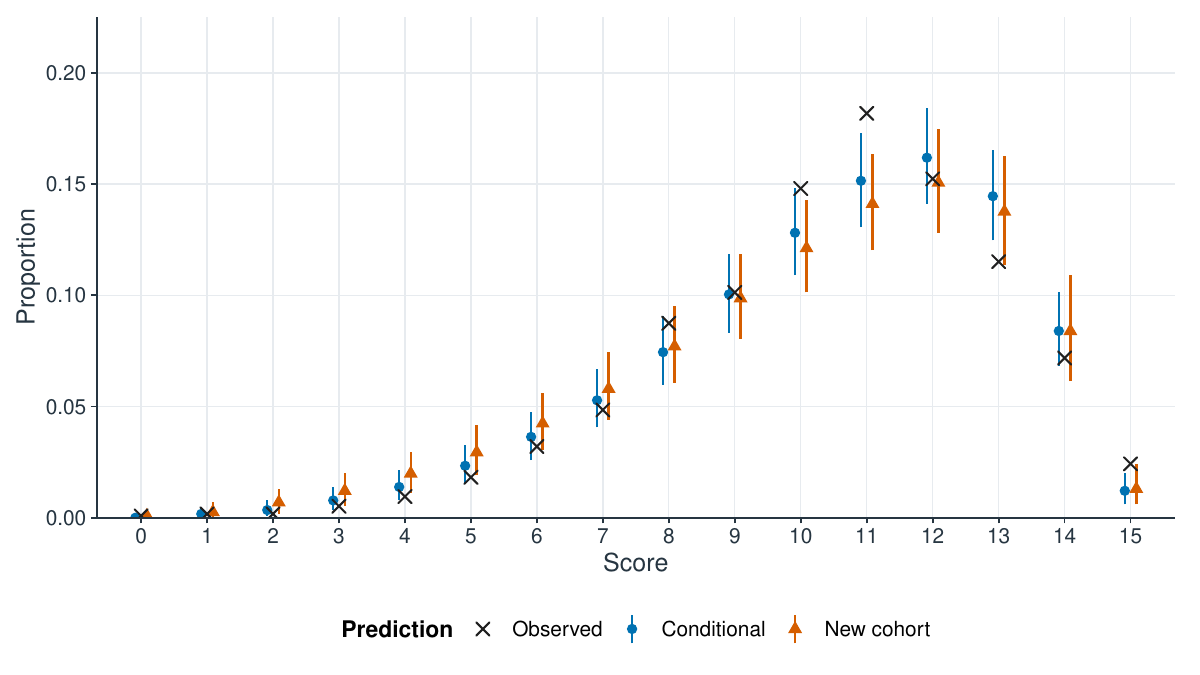}
\caption{Female}
\end{subfigure}

\smallskip

\begin{subfigure}{\textwidth}
\centering
\includegraphics[width=\linewidth]{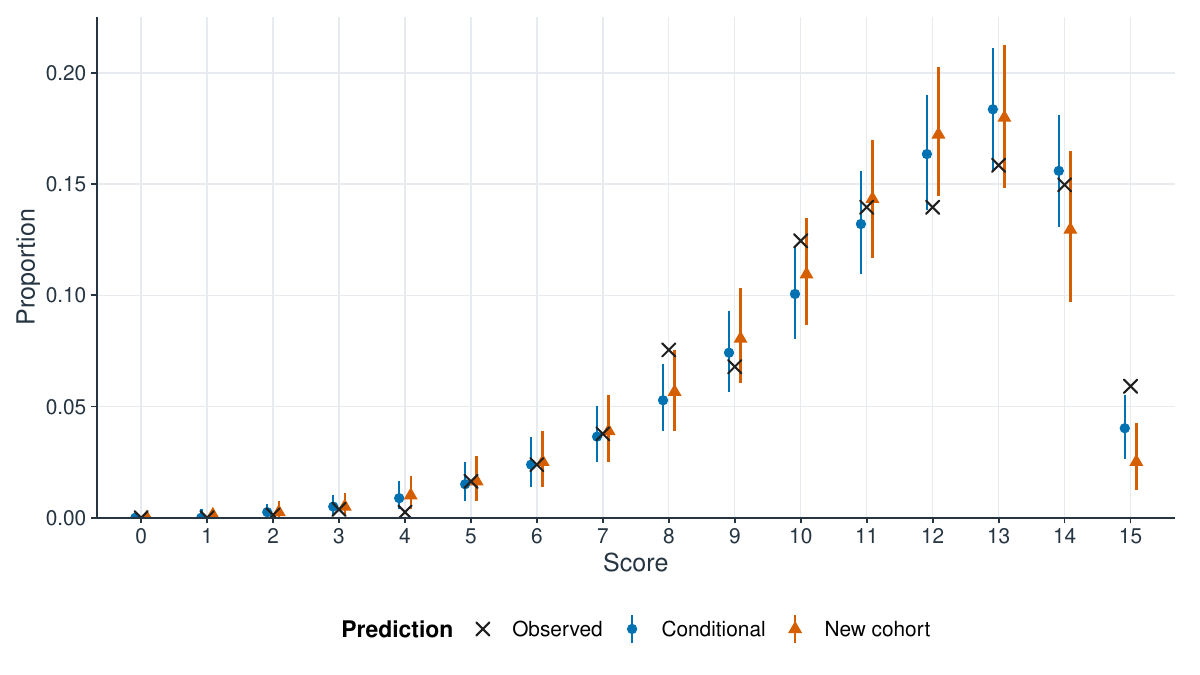}
\caption{Male}
\end{subfigure}
\caption{Posterior predictive BVRT score frequencies by sex for the model fitted at $\tau = 0.50$. Black crosses show observed proportions; blue circles and orange triangles show conditional and new-cohort predictive medians, respectively. Vertical lines indicate pointwise 95\% posterior predictive intervals. Proportions pool recorded assessments within each group. Prediction schemes and use of all post-warmup draws follow \suppfigref{fig:bvrt-frequencies-q50}.}
\label{fig:bvrt-frequencies-sex-q50}
\end{figure}
\clearpage

\begin{figure}[!htbp]
\centering
\includegraphics[width=\textwidth]{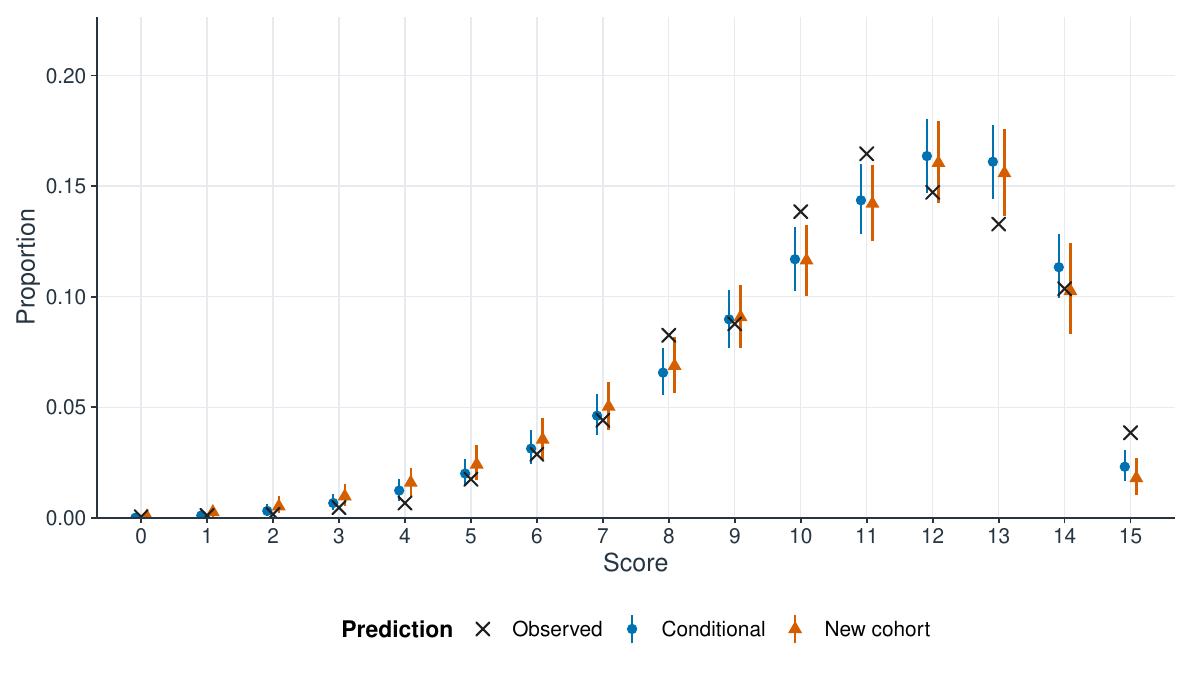}
\caption{Posterior predictive BVRT score frequencies overall for the model fitted at $\tau = 0.75$. Black crosses show observed proportions, blue circles show conditional predictive medians using the sampled effects of existing participants, and orange triangles show predictive medians using newly generated participant effects. Vertical lines indicate pointwise 95\% posterior predictive intervals. Proportions pool all recorded assessments. Both prediction schemes retain the observed covariates, assessment times, and pattern of recorded outcomes, with one shared effect vector per participant. All post-warmup draws are used.}
\label{fig:bvrt-frequencies-q75}
\end{figure}
\clearpage

\begin{figure}[p]
\centering
\captionsetup{font=small}
\begin{subfigure}{\textwidth}
\centering
\includegraphics[width=\linewidth]{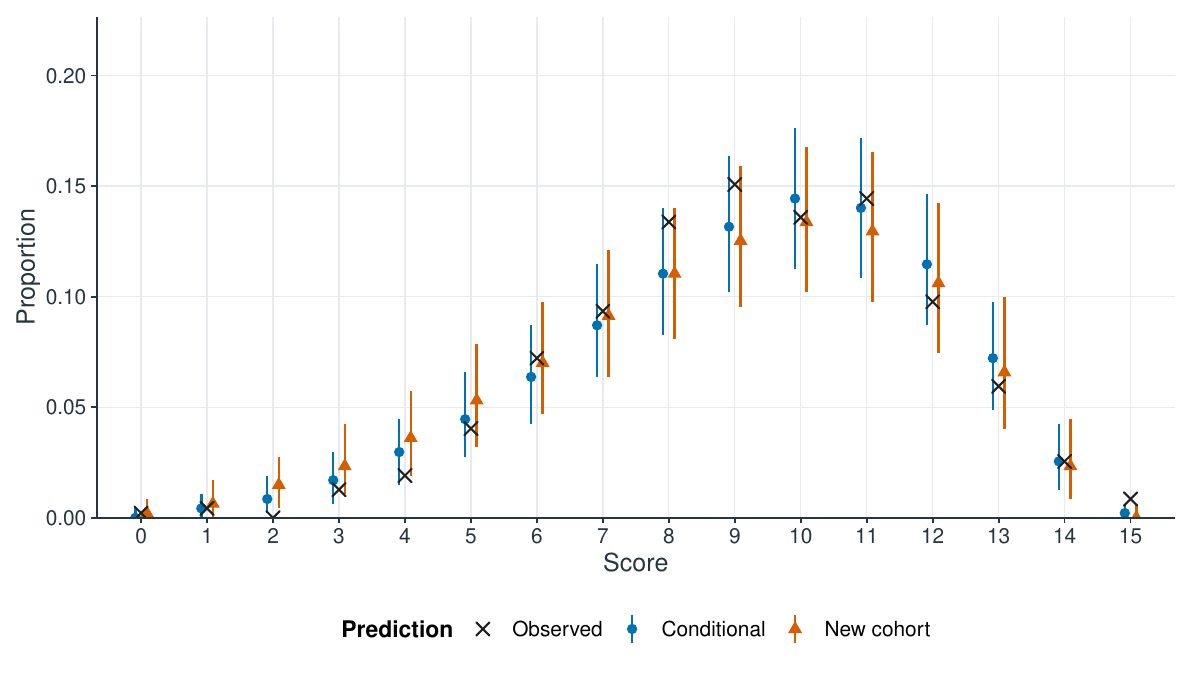}
\caption{No primary school graduation}
\end{subfigure}

\smallskip

\begin{subfigure}{\textwidth}
\centering
\includegraphics[width=\linewidth]{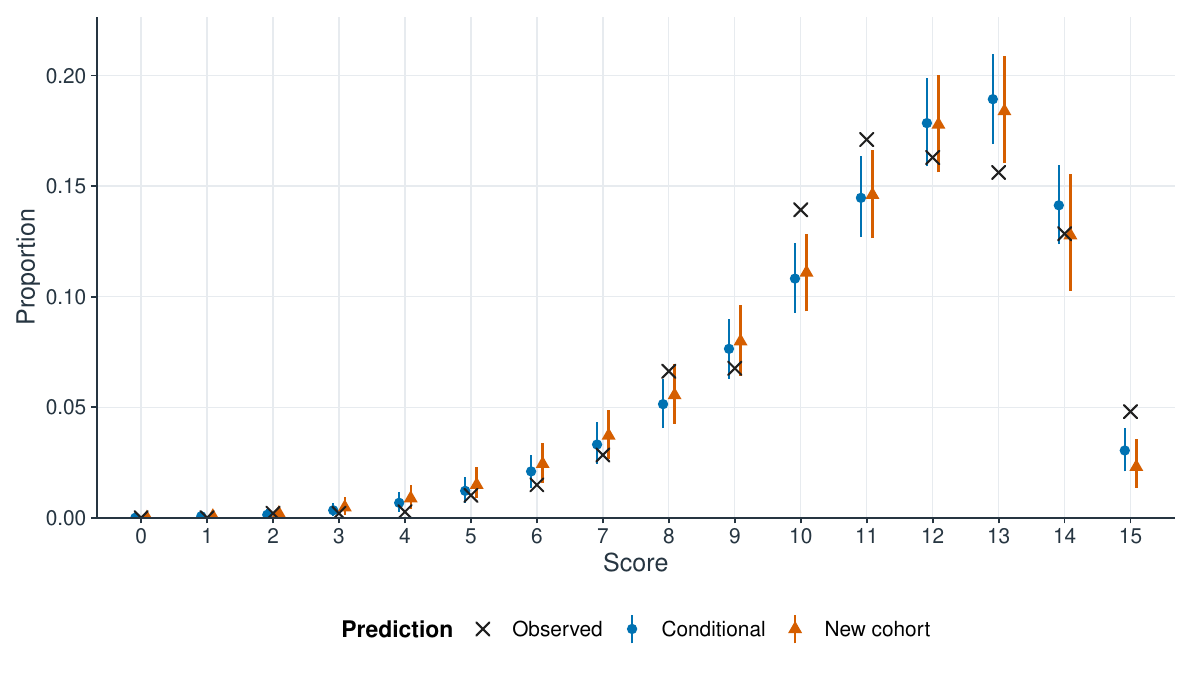}
\caption{Primary school graduates}
\end{subfigure}
\caption{Posterior predictive BVRT score frequencies by education for the model fitted at $\tau = 0.75$. Black crosses show observed proportions; blue circles and orange triangles show conditional and new-cohort predictive medians, respectively. Vertical lines indicate pointwise 95\% posterior predictive intervals. Proportions pool recorded assessments within each group. Prediction schemes and use of all post-warmup draws follow \suppfigref{fig:bvrt-frequencies-q75}.}
\label{fig:bvrt-frequencies-education-q75}
\end{figure}
\clearpage

\begin{figure}[p]
\centering
\captionsetup{font=small}
\begin{subfigure}{\textwidth}
\centering
\includegraphics[width=\linewidth]{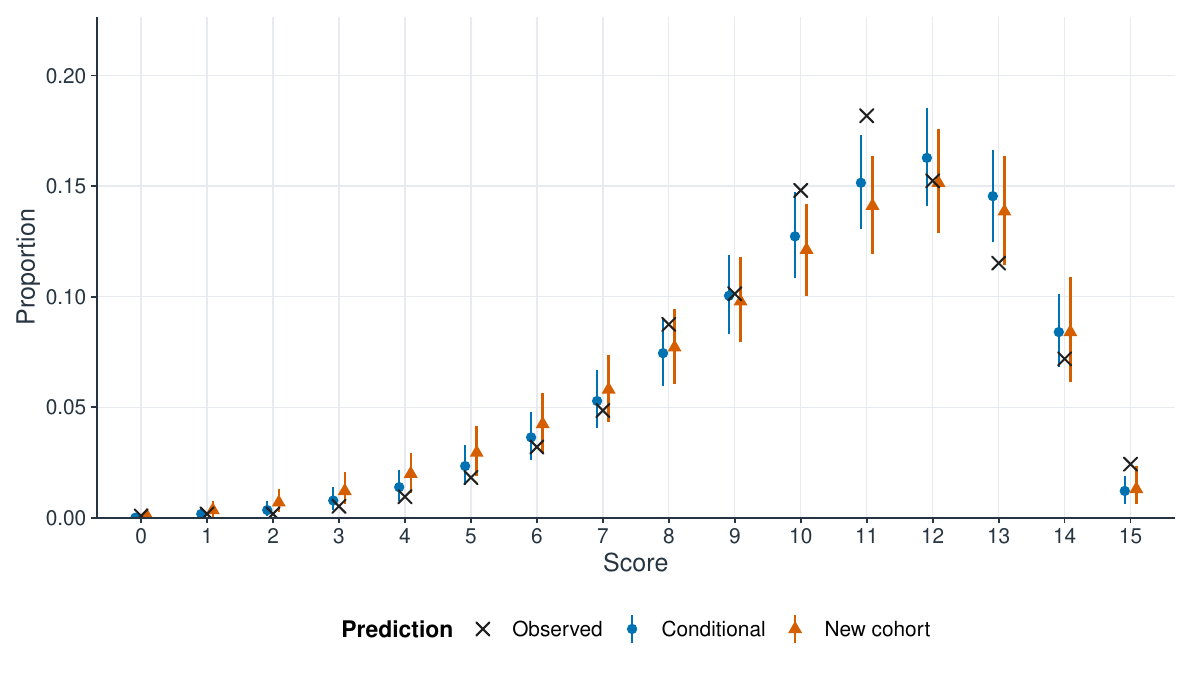}
\caption{Female}
\end{subfigure}

\smallskip

\begin{subfigure}{\textwidth}
\centering
\includegraphics[width=\linewidth]{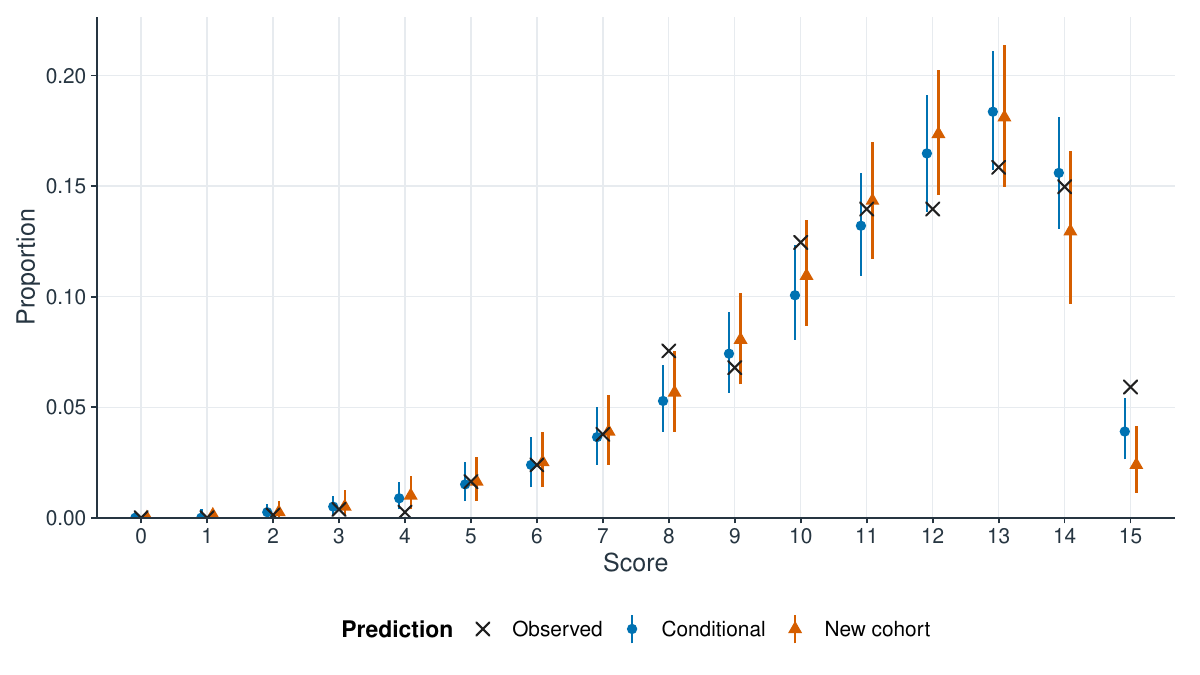}
\caption{Male}
\end{subfigure}
\caption{Posterior predictive BVRT score frequencies by sex for the model fitted at $\tau = 0.75$. Black crosses show observed proportions; blue circles and orange triangles show conditional and new-cohort predictive medians, respectively. Vertical lines indicate pointwise 95\% posterior predictive intervals. Proportions pool recorded assessments within each group. Prediction schemes and use of all post-warmup draws follow \suppfigref{fig:bvrt-frequencies-q75}.}
\label{fig:bvrt-frequencies-sex-q75}
\end{figure}
\clearpage

\begin{figure}[!htbp]
\centering
\includegraphics[width=\textwidth]{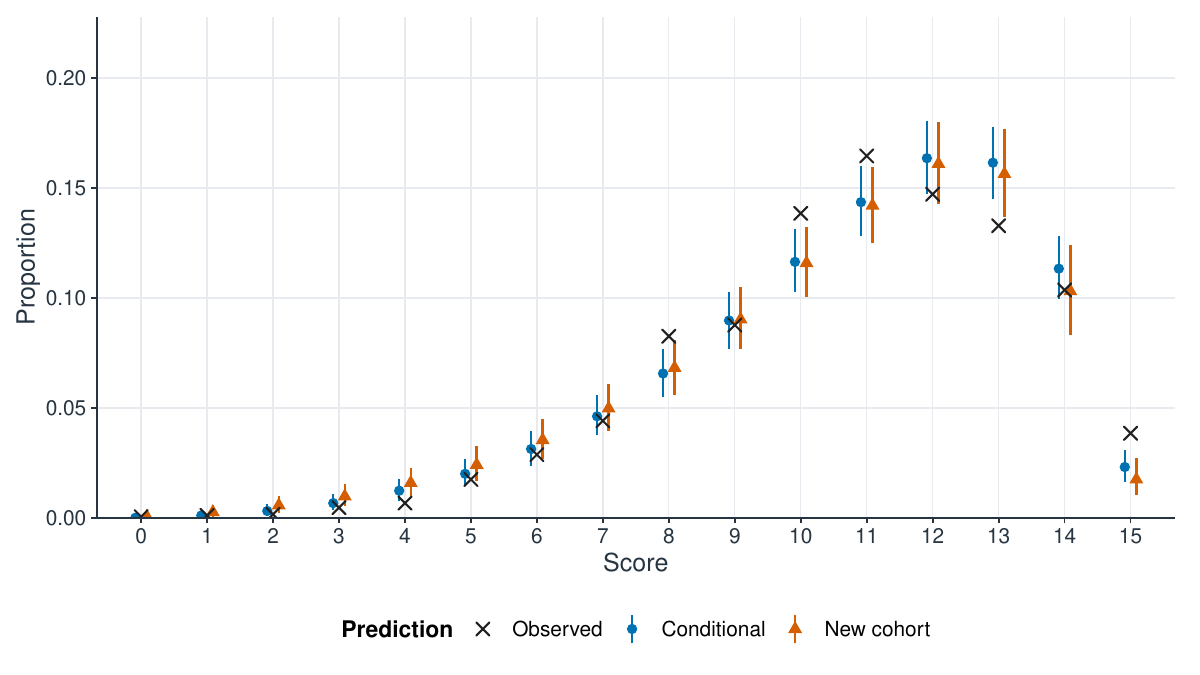}
\caption{Posterior predictive BVRT score frequencies overall for the model fitted at $\tau = 0.90$. Black crosses show observed proportions, blue circles show conditional predictive medians using the sampled effects of existing participants, and orange triangles show predictive medians using newly generated participant effects. Vertical lines indicate pointwise 95\% posterior predictive intervals. Proportions pool all recorded assessments. Both prediction schemes retain the observed covariates, assessment times, and pattern of recorded outcomes, with one shared effect vector per participant. All post-warmup draws are used.}
\label{fig:bvrt-frequencies-q90}
\end{figure}
\clearpage

\begin{figure}[p]
\centering
\captionsetup{font=small}
\begin{subfigure}{\textwidth}
\centering
\includegraphics[width=\linewidth]{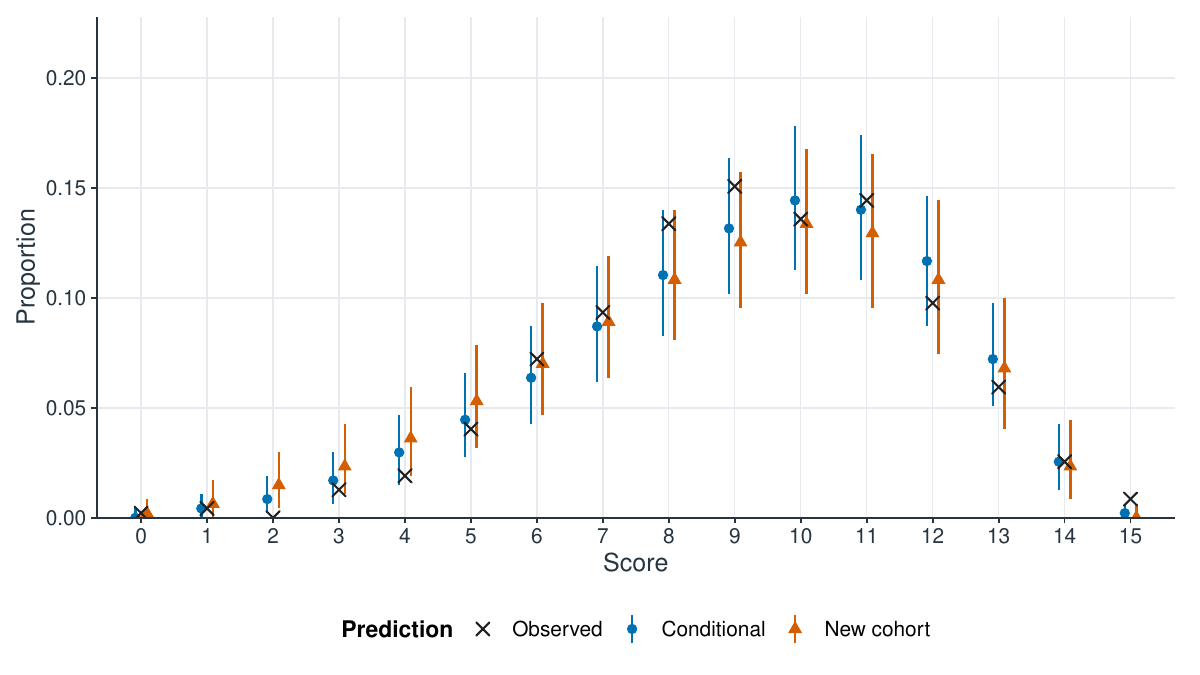}
\caption{No primary school graduation}
\end{subfigure}

\smallskip

\begin{subfigure}{\textwidth}
\centering
\includegraphics[width=\linewidth]{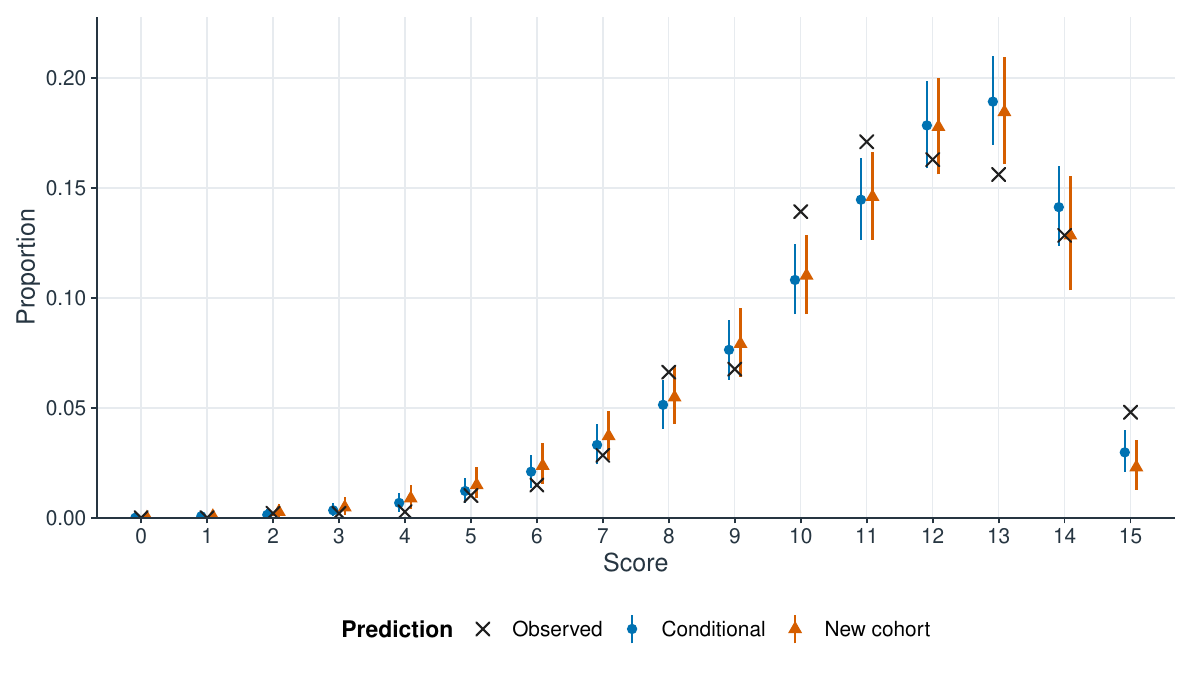}
\caption{Primary school graduates}
\end{subfigure}
\caption{Posterior predictive BVRT score frequencies by education for the model fitted at $\tau = 0.90$. Black crosses show observed proportions; blue circles and orange triangles show conditional and new-cohort predictive medians, respectively. Vertical lines indicate pointwise 95\% posterior predictive intervals. Proportions pool recorded assessments within each group. Prediction schemes and use of all post-warmup draws follow \suppfigref{fig:bvrt-frequencies-q90}.}
\label{fig:bvrt-frequencies-education-q90}
\end{figure}
\clearpage

\begin{figure}[p]
\centering
\captionsetup{font=small}
\begin{subfigure}{\textwidth}
\centering
\includegraphics[width=\linewidth]{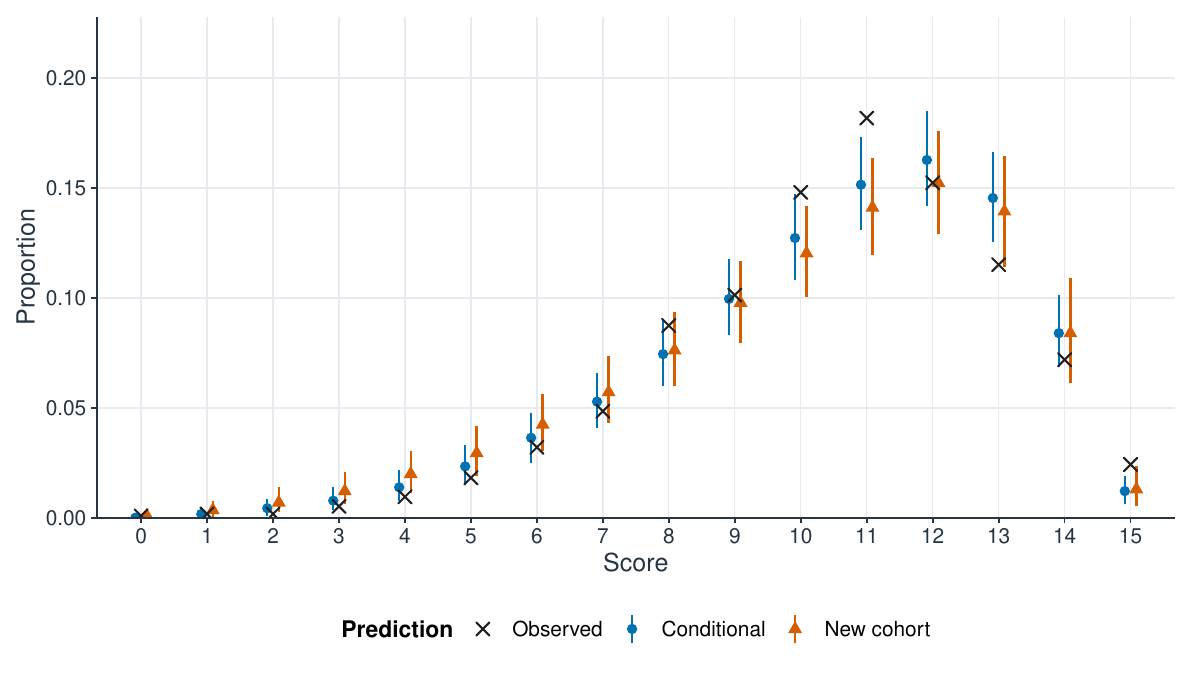}
\caption{Female}
\end{subfigure}

\smallskip

\begin{subfigure}{\textwidth}
\centering
\includegraphics[width=\linewidth]{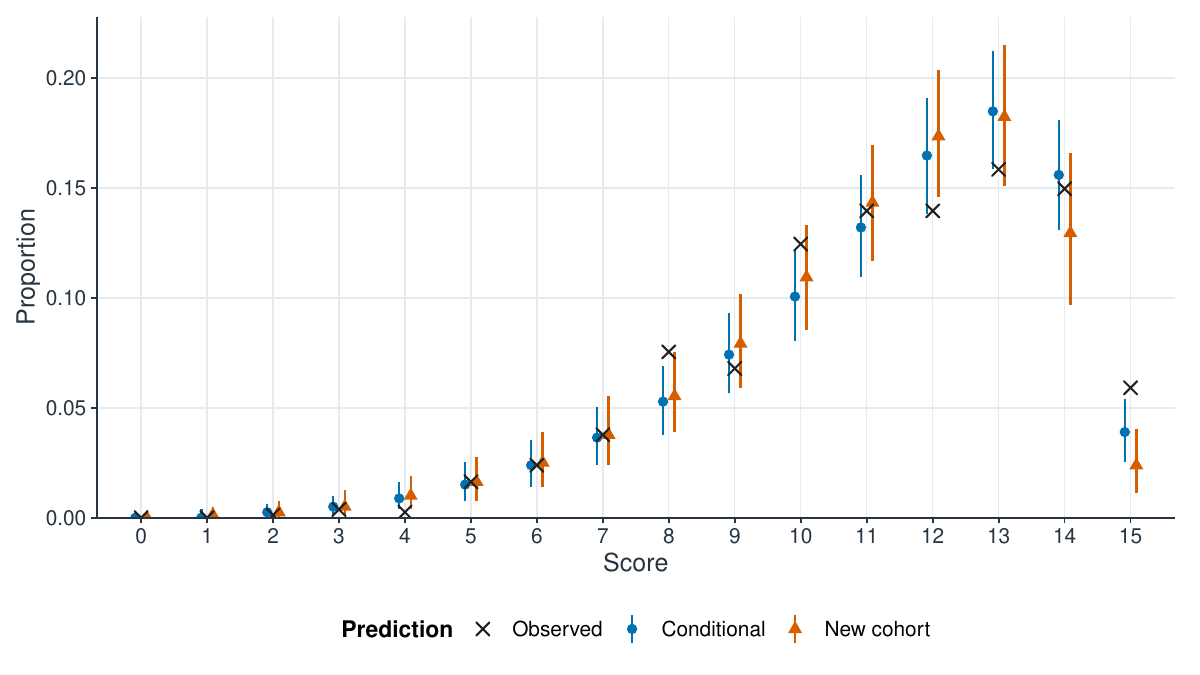}
\caption{Male}
\end{subfigure}
\caption{Posterior predictive BVRT score frequencies by sex for the model fitted at $\tau = 0.90$. Black crosses show observed proportions; blue circles and orange triangles show conditional and new-cohort predictive medians, respectively. Vertical lines indicate pointwise 95\% posterior predictive intervals. Proportions pool recorded assessments within each group. Prediction schemes and use of all post-warmup draws follow \suppfigref{fig:bvrt-frequencies-q90}.}
\label{fig:bvrt-frequencies-sex-q90}
\end{figure}
\clearpage

\FloatBarrier

\end{document}